# Detector R&D needs for the next generation $e^+e^-$ collider

A. Apresyan[i], M. Artuso[v], J. Brau[q], H. Chen[d], M. Demarteau[p], Z. Demiragli[c], S. Eno[n], J. Gonski[h], P. Grannis[u], H. Gray[e,m], O. Gutsche[i], C. Haber[m], M. Hohlmann[j], J. Hirschauer[i], G. Iakovidis[d], K. Jakobs[a], A.J. Lankford[g], C. Pena[i], S. Rajagopalan[d], J. Strube[r,q], C. Tully[s], C. Vernieri[t], A. White[w], G.W. Wilson[l], S. Xie[f], Z. Ye[k], J. Zhang[b], B. Zhou[o]

[a]Albert-Ludwigs-Universitat, Freiburg, [b]Argonne National Laboratory, [c]Boston University, [d]Brookhaven National Laboratory, [e]University of California, Berkeley, [f]California Institute of Technology, Pasadena, [g]University of California, Irvine, [h]Columbia University, New York, [i]Fermi National Accelerator Laboratory, [j]Florida Institute of Technology, Melbourne, [k]University of Illinois at Chicago, [l]University of Kansas, Lawrence, [m]Lawrence Berkeley National Laboratory, [n]University of Maryland, College Park, [o]University of Michigan, Ann Arbor, [p]Oak Ridge National Laboratory, [q]University of Oregon, Eugene, [r]Pacific Northwest National Laboratory, [s]Princeton University, [t]SLAC National Accelerator Laboratory, [u]Stony Brook University, [v]Syracuse University, [w]University of Texas at Arlington.

## Abstract

The 2021 Snowmass Energy Frontier panel wrote in its final report "*The realization of a Higgs factory will require an immediate, vigorous and targeted detector R&D program*". Both linear and circular $e^+e^-$ collider efforts have developed a conceptual design for their detectors and are aggressively pursuing a path to formalize these detector concepts. The U.S. has world-class expertise in particle detectors, and is eager to play a leading role in the next generation $e^+e^-$ collider, currently slated to become operational in the 2040s. It is urgent that the U.S. organize its efforts to provide leadership and make significant contributions in detector R&D. These investments are necessary to build and retain the U.S. expertise in detector R&D and future projects, enable significant contributions during the construction phase and maintain its leadership in the Energy Frontier regardless of the choice of the collider project. In this document, we discuss areas where the U.S. can and must play a leading role in the conceptual design and R&D for detectors for $e^+e^-$ colliders.

# Contents

## 1. Overview

"**Use the Higgs boson as a new tool for discovery**" was identified as one of the five compelling science drivers by the 2014 P5 committee [1]. Following the discovery of the Higgs boson [2, 3], the ATLAS and CMS experiments at the Large Hadron Collider (LHC) have made significant progress in quantifying its properties and will continue to do so during the High Luminosity LHC (HL-LHC) phase. The P5 committee went on to state that "*An $e^+e^-$ collider can provide the next outstanding opportunity to investigate the properties of Higgs in detail*", as it would greatly extend the sensitivity of the Higgs boson interaction with the Standard Model (SM) particles and with other new physics. Model-independent measurements of the Higgs coupling to some SM particles to sub-percent precision would allow a stringent test of the SM and could unveil small deviations, if any, from SM predictions. In addition to the rich program offered by an $e^+e^-$ collider using the Higgs boson as a tool, the large integrated luminosity accumulated at the Z-pole enables high-precision electroweak measurements and an ambitious flavor-physics program. It opens the door to the study of SM particles with unprecedented precision and observation of rare processes beyond SM expectations. In addition, operations at higher $WW$ and $t\bar{t}$ thresholds will further enhance sensitivity to new physics and provide a measurement of the Higgs self coupling.

A recent paper on Higgs Factory Considerations [4] submitted to Snowmass 2021 identified seven fundamental questions that can be addressed at a lepton collider which can operate in the energy range from the Z-pole to the TeV region. These include:

- Precision measurement of Higgs couplings to SM fermions and gauge bosons
- Measurement of Higgs self-couplings
- Sensitivity to rare or non-SM Higgs decays
- Discovery potential for new non-SM physics
- Ability to directly measure top electroweak and Yukawa couplings
- Sensitivity to new physics through precision measurement of loop effects
- Ability to improve precision of the strong coupling constant

In its 2020 report [5], the European Strategy for Particle Physics strongly endorsed the need for an $e^+e^-$ collider, stating "*An electron-positron Higgs Factory is the highest priority next collider*".

The HL-LHC program is expected [6] to end in the early 2040s after collecting 3000 fb$^{-1}$ of pp collisions at $\sqrt{s}$ ∼14TeV. As it will become evident in this document, now is the time to begin planning for the next-generation collider that will succeed and complement the physics offered by HL-LHC. The $e^+e^-$ colliders are the most technologically advanced, providing the most promising opportunity to follow the HL-LHC program. There are several proposals that are being considered by the international community, including an International Linear Collider (ILC) in Japan and a Future Circular Collider (FCC) at CERN. Other proposals, such as the Cool Copper Collider ($C^3$) are also under discussion in the U.S. In a technologically limited schedule, both circular and linear e⁺e⁻ colliders are ideally positioned to begin operations in the 2040s, as both are based on well-developed accelerator technology. Operations of an e⁺e⁻ collider would seamlessly follow the conclusion of the HL-LHC program and match well with both physics goals and community needs.

Figure 1 shows the proposed timeline for the deployment of FCC and ILC as presented by the respective host laboratory Directors [6] [7]. It shows the approval for the construction of the respective collider to be made around 2028. In both cases, civil construction would begin around 2030 if approved. The ILC projects a success-oriented schedule that aims to complete construction and installation in the late 2030s and begin physics running in 2040. The FCC schedule is aligned, as required, with the HL-LHC schedule and projects completion of construction and installation in mid-2040s.

The detector concepts and designs are largely common for both the Linear and Circular colliders, as will be evident in this document. Software and computing efforts are also synergistic, both building on a common suite of software tools and framework. It is necessary to pursue and study multiple detector technologies that would provide foundational detector concepts for one or more experiments at any of the e⁺e⁻ colliders. Given the high degree of overlap in detector concepts and the aligned timeline for pursuing the R&D, the U.S. circular and linear collider communities (FCC, ILC and $C^3$) have developed this coherent and coordinated funding proposal to the P5 committee for their consideration.

This document focuses on the proposed near-term U.S. participation in a targeted detector R&D and software development program that is required to enable U.S. physicists to take on leadership roles in the next generation e⁺e⁻ colliders.

*1.1. Detector R&D timeline and strategy*

The approval of the next generation e⁺e⁻ collider and the start of operations serve as the reference points to plan for detector R&D and construction. Much like the LHC, emerging experimental collaborations will develop their Technical Design Reports (TDR) that define the detector building blocks and subsequently seek its endorsement soon after the formal approval of the e⁺e⁻ collider program. Hence, the next several years, leading to the approval of the e⁺e⁻ collider, is a critical phase to pursue a targeted detector R&D program to identify technologies and prepare the groundwork to influence the detector design concepts for each experiment. This is indicated as phase (1) in the timeline shown in Figure 1. Following the approval of the e⁺e⁻ collider, experimental collaborations begin to coalesce formally and begin to document the chosen detector concepts in their TDRs and seek its approval. The preparatory phase leading to the approval of the TDR, including continued R&D to to finalize the broad detector design concepts, is indicated as phase (2) in Figure 1. Following the TDRs, the detector construction phase, which includes prototyping, pre-production, and production of the various detector elements and its subsequent integration and installation, typically takes 10 - 12 years, following the experience gained at other large experiments including the HL-LHC upgrades. This is indicated as phase (3) in Figure 1. Phase (4) reflects the commissioning with beam and subsequent physics running.

By combining strengths and exploiting the synergies between the circular and linear collider communities, U.S. physicists can coherently pursue the critical R&D required for the next generation colliders in a cost-effective manner. Uncertainties in the accelerator technology for future $e^+e^-$ colliders makes the need for having a cohesive approach to detector R&D even more vital as it ensures U.S. to be prepared regardless of which collider option is ultimately chosen.

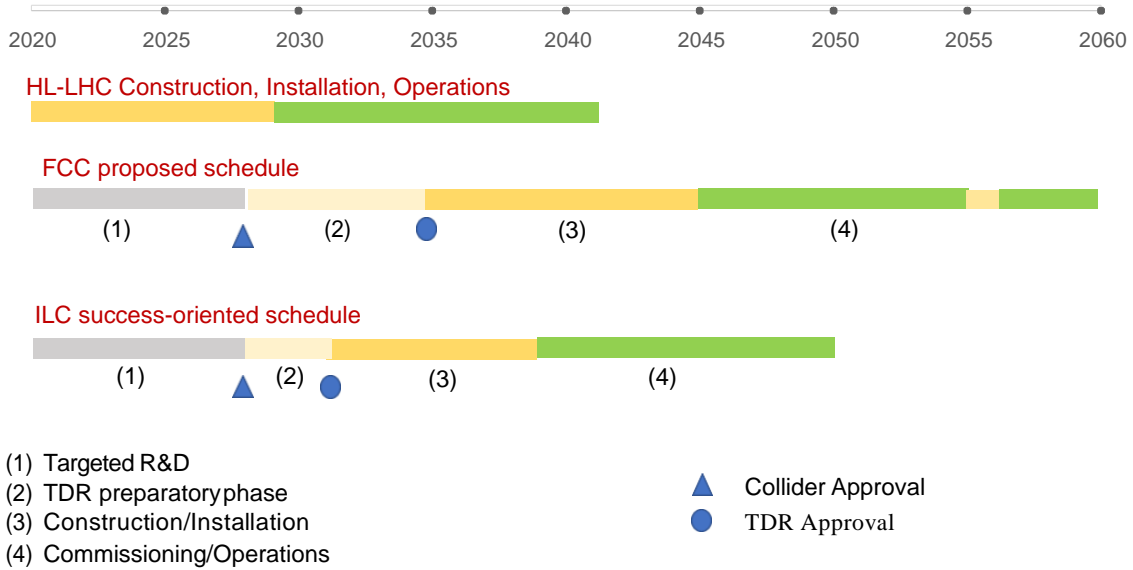

Figure 1: Timelines for HL-LHC, FCC and ILC collider projects, as proposed by the respective laboratory managements [6] [7], showing the major phases of R&D, construction, and operations.

Efforts to build on the technologies pursued by the HL-LHC upgrades and collaboration with other major U.S. projects such as the EIC are already underway. Collaboration and complementarity of U.S. led detector R&D programs with other ongoing international efforts are necessary and these communication channels have recently opened. While this document expresses the interests and expertise of the U.S. groups, the coming months will further focus these expression of interests following the discussions and negotiations with international partners as well as exploiting synergies with other U.S. HEP and NP (Nuclear Physics) groups.

Engagement of U.S. physicists in targeted detector R&D will not only allow the U.S. to exploit its expertise and interests to influence the detector concepts but will also allow the U.S. to assume major roles during the construction phase. These investments will enable the U.S. to build international partnerships, maintain leadership in the Energy Frontier, and exploit the physics that such colliders have to offer.

The timescales also define the critical decision points as laid out in the DOE 413.3b Project Management process as well as the major stages of the NSF MREFC process. A DOE Mission need (CD-0) can be expected soon after the approval of the $e^+e^-$ collider and a CD-1 following the completion of the TDRs and a broad agreement on the scope of the U.S.

contribution during the construction phase. The NSF MREFC proposal would be pursued in parallel with the goal of seeking its approval around the same time as the DOE CD-1 approval, thus allowing the construction phase to move forward in tandem.

*1.2. Organization*

A U.S.-wide coordination body was formed to plan and develop the scope of the detector R&D efforts targeting future $e^+e^-$ colliders. Following the recent road map published by the European Committee for Future Accelerators (ECFA) in 2020 [8] and a similar study in the U.S. leading to the Basic Research Needs road map [9], a number of international R&D collaborations grouped in technological themes are in the process of being formed. The U.S. $e^+e^-$ detector coordination chose to align itself consistently with these technological panels to provide efficient communication and partnership with other entities and exploit the synergies. The coordination group is organized along the following themes:

- Solid State Devices focusing on inner tracker detector concepts
- Calorimeter, including noble liquid-, silicon-, crystal- and scintillation-based readout calorimeters
- Gaseous detectors, focusing on muon spectrometer and gaseous inner tracker
- Particle ID, focusing on specialized detectors to support particle identification
- ASICs/Electronics, focusing on providing developmental support across all technological groups
- Trigger/DAQ: focusing on smart triggering and data readout
- Quantum Devices: focusing on potential integration of novel quantum technologies into detector design
- Software/Computing, focusing on providing the required software infrastructure and tools for simulation, data processing, and detector design/optimization

The LHC coil technology used in the production of solenoid magnets are no longer supported by the industry. Hence, investments are critically required to find alternative solution for the next generation experiments. While no U.S. groups have currently expressed interest in pursuing this study, and therefore is not reflected in the above list, the U.S. with its strong record in magnet technology may well be able to contribute to this effort.

Each of the above-mentioned groups above were charged with engaging the U.S. community, exploiting the U.S. strengths to define the scope of the R&D program. Input for this proposal, defined in the subsequent sections, have been driven by these technological groups. Included in their responsibility was to collaborate with other entities across HEP and NP and exploit synergies to develop a focused, coherent and cost-effective program.

Note that this document reflects the current U.S. interests. As collaborations form for these international efforts, the U.S. responsibilities will naturally evolve and adapt.

*1.3. Near Term R&D needs*

The detector R&D needs for the future $e^+e^-$ colliders have been developed as a bottom-up community driven exercise. Each group's coordinators, jointly responsible for addressing the needs for both circular and linear colliders, were charged with engaging the community and gauging their interests and expertise. A list of R&D topics for each group was documented, based on the long-standing expertise in the U.S. and the interests of the community. The R&D topics were then prioritized into three categories: High, Medium, Low following the prioritization guidelines laid out by ECFA detector road map:

- **High**: R&D that is critical to achieving the physics requirements
- **Medium**: R&D that is important to achieve the physics objectives, provide more cost-effective solutions and reduce complexity.
- **Low**: R&D that can potentially further enhance the physics reach.

The following sections document the R&D efforts required in each technology group over the next decade required to meet the objectives of the TDR. The scope and justification of each R&D topic, and a timescale reflected through high level milestones are defined.

The list of R&D topics represents the current interests of the U.S. community and where domestic resources/expertise are available and can be exploited. Synergies with other projects, including international efforts are identified where possible. Negotiations with international partners are ongoing to collaborate on common efforts and identify areas that are unique to U.S. These efforts will further focus the proposed R&D efforts to ensure complementarity and a cost-effective strategy for the U.S. program.

Figure 2, and 3 shows the summary of the R&D requests for each group, that are further documented in subsequent sections. The priority for each of the listed R&D effort and the key parameters that they are intended to address are also shown.

*1.4. Conclusion*

The 2021 Snowmass Energy Frontier panel wrote in its final report "*The realization of a Higgs factory will require an immediate, vigorous and targeted detector R&D program.*". Both Linear and Circular collider efforts have developed a conceptual design for their detectors and are aggressively pursuing a path to formalize these detector concepts. It is urgent that the U.S. organize its efforts should it choose to provide leadership and make significant contributions to the future experiments. These investments are necessary to build and retain the U.S. expertise in detector R&D and future projects and maintain its leadership in the Energy Frontier regardless of the choice of the collider project. We urge P5 to recommend an R&D program for detectors that will sustain the US leadership in a global Energy Frontier research.

## 2. Solid State Tracking

*2.1. Challenges for Solid State tracking detectors*

Precision inner tracking, covering a barrel and forward/backward region, is a key feature of any high energy electron-positron collider detector. Various detector schemes have been discussed which range from a full silicon system (vertexer + tracker), to a hybrid design with the silicon vertexer at relatively small radius combined with a large, low mass TPC or drift chamber. To support particle identification by time-of-flight, an outer silicon "wrapper" is also included beyond the tracker. The wrapper would feature fast (~10 ps) timing detectors with the segmenation set by occupancy requirements, rather than precision. The technologies which could meet the requirements of solid state tracking, and timing, at a future electron-positron collider have also been discussed extensively in the DOE Basic Research Needs Study (2020), the ECFA Detector R&D Roadmap (2021), and the Snowmass Report of the Instrumentation Frontier (2022). Detailed tables of requirements for these various options can be found in these documents. One interesting conclusion which emerges from these recent studies, is the near identical requirements, for solid state tracking, at a future electron-positron collider, and the now planned Electron-Ion Collider (EIC). With the EIC on a shorter timescale than the electron-positron machine, we can see important opportunities for collaboration across the DOE Office of Science, and for intermediate scale deployments of some of the new technologies which will be considered.

| Section | R&D Topic | Priority | Key Targets |
|---|---|---|---|
| 2 | **Solid State Tracking** | | |
| 2.4.1 | 10 ps Timing from LGADS for Particle ID | medium | resolution=10 ps or better |
| 2.4.2 | Further Development of Sensor Expertise in the United States | high | MAPS or alternatives |
| 2.4.3 | System Integration for Low Mass High Precision Trackers | high | low mass "stave" structures at 1% of a radiation length |
| 2.4.4 | Development of Low Mass Support and Cooling Structures | high | 1% of a radiation length for outer tracking |
| 2.4.5 | High Efficiency Powering and Readout Schemes | high | Pulsed power and reducing power needs |
| 3 | **Muons Detectors and Gaseous Detectors** | | |
| 3.4.1 | Large-Area Muon Detectors with Fast Readout and High Precision | high | eco-friendly gases, 80 um res. bending plane, O(100 ps) timing electronics, robustness and redundancy |
| 3.4.2 | US Based R&D Facility for MPGD with Nuclear Physics Community | high | MPGD component production in US at JLAB |
| 3.4.3 | Low Mass Gaseous Detectors for Outer Region Main Tracking | high | drift chamber, straw tubes, and MPGDs competitive with outer Silicon layers in mass, resolution, and pattern recognition at lower cost |
| 3.4.4 | Services and Infrastructure for Gaseous Detectors | medium | HV, gas, and alignment |
| 4 | **Calorimetry** | | |
| 4.1 | CMOS MAPS EM Calorimetry | high | EM energy resolution < 3% and two EM shower separation < 2 mm at 50 GeV with nsec timing and < 1 uW/pixel |
| 4.2 | Noble Liquid Calorimetry | high | Superior SNR with cold electronics, fine segmentation for PFA |
| 4.3 | Hybrid Dual Readout Optical Calorimetry | high | compensation/particle flow |
| 4.4 | Scintillator Tiles with SiPMs | high | hadron calorimetry with 3-4% jet energy resolution, efficient track following for PFA |
| 4.5 | RPC Readout Digial Calorimeters | low | low cost/large area |
| 5 | **Particle ID** | | |
| 5.3.1 | LGAD Time of Flight | high | < 10 ps system performance |
| 5.3.2 | LAPPD Time of Flight | high | < 10 ps system performance |

Figure 2: Summary of prioritized R&D activities and key R&D targets

| Section | R&D Topic | Priority | Key Targets |
|---|---|---|---|
| 6 | **Readout Systems and ASICs*** | | |
| 6.3.1 | AI/ML in ASICs | high | Successful design & fab of ML-based readout prototype |
| 6.3.2 | Monolithic Sensor ASICs | high | Successful design & fab of monolithic sensor ASIC |
| 6.3.3 | High Performance ASICs for 4D/5D Systems | high | < 10 ps timing resolution |
| 6.3.4 | IP Blocks for 28 nm Technology | high | Successful design & fab of 28nm prototype |
| 6.3.5 | 3D/Hybrid Integration | high | Low mass high performance options |
| 6.3.6 | ASICs for Silicon Photonics | high | >50 Gbps readout |
| 6.3.7 | Increased Data Density | medium | Cope with higher data rates without sacrificing performance |
| 6.3.8 | Emerging Technologies | medium | Exploit latest technologies to improve performance |
| 6.3.9 | Extreme Environments | low | Ensure ASIC performance in varying environments |
| 7 | **Trigger and Data Acquisition** | | |
| 7.3.1 | Applications of Machine Learning to TDAQ | high | AI/ML/neuromorphic processing at us level |
| 7.3.2 | Achieving High Precision Timing Distribution | medium | 25 ps synchronization across varying scales |
| 7.3.3 | Integration of Modern Computing Hardware | high | heterogenous and streaming architectures |
| 7.3.4 | Improving Data Link Performance and Alternatives | medium | assess COTS at 40-400 Gbps |
| 8 | **Software and Computing** | | |
| 8.3 | Core Software | high | Common core framework; detector simulation development |
| 8.4 | Infrastructure | high | HTC/HPC facilities, adoption of community solutions |
| 8.5 | Physics Software including AI/ML | medium | N/A |
| 8.6 | Coordination | medium | N/A |
| 8.7 | User Support | low | Support for collaboration |
| 9 | **Quantum Sensors** | | |
| 9.4.1 | Superconducting Nanowire Sensors | low | 1000 ch/sensor, 5x5 cm^2, < 10 ps; rad. hard, hight-TC materials |
| 9.4.2 | Low Dimensional Materials | low | photocathodes/scintillators |

Figure 3: Summary of prioritized R&D activities and key R&D targets

Broadly stated, the requirements and challenges for solid state tracking at a Future electron-positron Collider Detector (FCD) are as follows.

- **Precision:** While present generation trackers function at the ~ 10 micron scale, these future trackers will require greater precision, typically 3 (6) microns for the vertex (tracking) layers. This pushes us to greater circuit densities and segmentation.

- **Mass:** These trackers require minimal mass leading to novel support structures, cooling strategies, and sensor configurations. Following the pioneering work of the heavy ion collider communities, future trackers will rely heavily on thinned monolithic active pixel sensors (MAPS), or other novel sensor structures. The requirements for vertex layers approach an equivalent thickness of 50 microns of silicon, implying a fully active and self supporting structure. For the outer tracking layers the requirement relaxes to 1% of a radiation length, still challenging but allowing for additional support and services.

- **Power:** Highly efficient powering schemes will be required as part of the mass reduction in services and cooling. Generally these will evolve from the present generation of serial and DC-DC conversion based systems, but may also depend on power pulsing for low collision rate environments. These power limits range from 20 to 100 mW/cm$^2$.

- **Scale:** Aspects of the before-mentioned technologies have already been applied, but at a much smaller scale than will be required at these future colliders. Consequently, the community will have to increasingly adopt industrial sourcing, and highly optimized assembly and test processes. Some of this is already being utilized for the HL-LHC upgrades albeit for structures with more modest mass and cooling requirements.

- **Timing:** Fast (several 10's of picoseconds) timing is being prepared for the HL-LHC upgrades, and is planned for the EIC as well. These are already fairly extensive systems. In the case of the HL-LHC, timing allows us to associate tracks to specific vertices in the presence of multiple interactions. In the case of the EIC, timing is used for particle identification. At a future electron-positron collider, timing is only envisioned, at present, for particle identification, using the outer silicon wrapper layer. The timing technology now being deployed for the HL-LHC may not scale directly to the needs of the the future electron-positron collider, for which time-of-flight requirements push us to all the way to 10 ps resolution, and potentially beyond. Consequently the timing performance will have to be improved and the sensor/readout technology may also evolve towards lower mass,

*2.2. Relevant US expertise in Solid State tracking detectors*

While the technologies required for a future electron-positron collider present a variety of challenges, a strong community already exists, in the United States, with experience and motivation to address these.

- The LHC and HL-LHC community, and the earlier Tevatron collider community, have extensive experience in silicon strip and silicon pixel detector development and large scale implementation. University and national laboratory groups have experienced teams of ASIC designers on staff. While the development of MAPS is an even more specialized skill set, the US ASIC community has led the development of pixel circuits as well as data acquisition, control, and power management ASICs which reside on detectors.

- The pioneering work on deploying low mass MAPS, spanning electronic, mechanical, and thermal aspects, occurred at the Brookhaven Relativistic Heavy Ion Collider for the STAR Heavy Flavor Tracker. Elements of this community are now focused on MAPS tracking for the EIC.

- US groups participate in B-factories, up to the present day, and there is considerable solid state tracking experience there.

- US groups play a major role in the development of fast timing. This covers the fundamental work on the Low Gain Avalanche Detector (LGAD), over the last ten years, culminating in the corresponding ATLAS and CMS forward timing layers. CMS also is deploying fast Silicon Photomultiplier (SiPM) devices in their barrel region. Also importantly, these large deployments mean we will gain important experience in fast timing systems including calibration, synchronization, and so forth.

- US groups have played leading roles in the development of low mass, carbon based, support structures. University and national laboratory groups have specialized facilities for the design and processing of carbon structures and deep connections to the industries which provide these materials.

US groups should play a major role in the design and fabrication of a detector of the future electron positron colliders. In determining an appropriate course, we must balance a number of factors. These would include existing expertise, impact, our interest in entirely new challenges, and of course the ambitions of our non-US colleagues. In any event, it would be hard to imagine that US groups would not contribute to inner tracking and/or timing in such a collaboration. Furthermore, it is natural to assume that our buy-in would be at least 20% of cost, to an off-shore project, and be coupled to significant technical contributions. Issues to consider include the following.

- MAPS is certainly a major enabling technology for the future collider tracker. US groups will need to ramp up their involvement in MAPS to have impact on an FCD. We will need to understand, how far from the present performance of MAPS we need to go to meet the needs of an FCD. We will want to identify particular MAPS challenges where the US groups could make important contributions.

- Low mass support structures, cooling, and power management are areas with already significant US expertise. There will likely be strong arguments to retain and grow this. These are crucial technologies for tracking at an FCD and represent significant opportunities for the US community,

- While the development of fast timing has been an international effort, certainly the US has played a huge role here. Consequently it would be natural for the US to lead, or take on major commitments here. The question however becomes the scale and level of importance of fast timing at an FCD. Current studies indicate that fast timing would be unnecessary in the inner layers. On the other hand, the large radius silicon wrapper layer needs to provide both a space point, with appropriate precision and low occupancy, and an unprecedented timing measurement, in support of particle identification. Thus the silicon wrapper could be a natural, and self-contained deliverable for the US to take responsibility for, both technically, and financially.

- There may be new or emerging ideas and/or technologies which could impact the design and performance of the FCD. This issue was highlighted in the DOE BRN report Priority Research Direction 19. Such ideas which could include new thin film sensors, non-silicon materials, Shockley-Ramo induction sensors, heterogeneous structures, and others, assuming they could converge on the necessary timescale.

These aspects of potential US involvement branch into a set of five R&D tasks which are described in Section 2.4 below. Each would be considered first by an initial study to further understand the opportunities, costs, and implications. The outcomes lead to an informed and technically prioritized R&D plan for the US, for the next 10 years based upon the baseline schedule shown in Figure 1.

Among the charges to these studies are as follows.

- Determine what opportunities exist in MAPS for additional innovation required for an FCD. Determine which US groups would want to enter into this activity and collaborative opportunities with non-US groups. Understand what, if any, impediments there are to access the necessary foundry processes.

- Consider the deployment of a large low mass structure with appropriate cooling methods on a scale set by a) a standalone inner tracker, b) a complete silicon tracker, and c) a silicon wrapper layer (outside a drift chamber or TPC).

- What technology would we need to develop and deploy for a silicon wrapper layer including fast timing?

- What of any new or emerging technologies (other than MAPS) could offer performance gains for an FCD which would warrant the risk inherent in their development? Is there a critical mass in the US for any of these directions?

Following the studies, it will be natural to focus on one or more pilot projects. For example, we should plan to develop a large scalable thinned MAPS tracking stave (task 3). We would need to take into account electrical, mechanical, and thermal services. This would significantly inform the needs for a future production project. Similarly we should develop a scalable segment of the silicon wrapper timing layer (task 1) including services and supports in collaboration with the broader effort on particle ID and on front end electronics. It would also be natural to construct a global support structure "sector" demonstration prototype (task 4).

*2.3. US Solid State Tracking Institutions*

Institutions responding to a survey or expressing interest in solid state tracking detector development for future $e^+e^-$ colliders: Argonne National Laboratory, Brookhaven National Laboratory, California Institute of Technology, Duke University, Fermi National Accelerator Laboratory, Lawrence Berkeley National Laboratory, Oak Ridge National Laboratory, SLAC National Accelerator Laboratory, Stony Brook University, Purdue University, University of California Santa Cruz/SCIPP, University of Chicago, University of Illinois Chicago, University of Massachusetts Amherst, University of Oregon, University of Texas at Arlington, University of Washington

*2.4. List of Solid State detector R&D tasks*

*2.4.1. Solid State detector task 1: 10 ps timing from LGADs for Particle ID*

- **Title: 10 ps Timing Resolution Using LGADs for Particle Identification**
- **Duration:** 10 years
- **Priority:** medium
- **Justification:** Particle ID is an important aspect of the high statistics Z physics and the heavy flavor programs. Its importance to the higher energy phases is less critical, for this reason we assign it medium priority. The basic time resolution specification for time of flight, as part of an integrated particle ID system, depends upon the required K-$\pi$ separation, the momentum range, and the dimensions of the system. For example, the extension of a 5 $\sigma$ K-$\pi$ separation from up to 10 GeV, upwards to 20 GeV, requires a resolution of 10 ps at 2 meters. This time resolution is at or beyond the edge of that which is currently achieved and therefore sets a target scale for this R&D topic. Within Solid State tracking the focus will be on the sensor development. The companion Particle ID activity would focus on system aspects while the Readout Systems and ASICs activity would focus on the front end electronics, control, and data transmission aspects. This three pronged approach is reflected in the resource requirements of these three areas.
- **Milestones:**
    – Planning and consideration of options and targets Year 1
    – Demonstration of technology towards 10 ps Year 1-2

- Production of large-area sensors with uniform performance with 10 ps Year 3-4
- System design and system test Year 5-6
- Design for full scale production and final prototyping Year 7-10

- **Institutes:** SLAC, Fermilab, UC Santa Cruz, U Chicago, U Iowa, BNL, LBNL, Argonne, U Illinois Chicago

*2.4.2. Solid State detector task 2: Further Development of Sensor Expertise in the USA*

- **Title: Development of Low Mass High Precision Sensor Expertise in the USA**
- **Duration:** 10 years
- **Priority:** high
- **Justification:** MAPS is widely viewed as the enabling technology for a lightweight tracker at future electron positron colliders. There are also other emerging approaches which may also show promise. There is effort in the USA targeting MAPS for the Electron Ion Collider. The US should play a significant role here and we need to rapidly ramp US HEP effort in this area. This encompasses design expertise and vendor engagement. Access to appropriate foundry processes key to a production R&D in this area.
- **Milestones:**
  - Study group to determine the US scope and the deliverables for sensor prototyping and later involvement. Do we integrate with an existing effort or focus on a standalone project? Year 1
  - Acquire design expertise Year 1-2
  - First prototypes submission Year 3
  - Technology down select Year 3-5
  - Large scale prototyping and development of a production testing and packaging process Year 6-8
  - Final prototying and preparation for production scale orders and testing Year 8-10
- **Institutes:** SLAC, U Oregon, U Texas Arlington, BNL, LBNL, Argonne, Fermilab, UC Santa Cruz, U Chicago, Duke, Caltech

*2.4.3. Solid State detector task 3: System level integration aspects for a low mass high precision tracker*

- **Title: System Integration for Trackers**
- **Duration:** 10 years

- **Priority:** high

- **Justification:** While any sensor, thinned to ~50 microns, is inherently low mass, the rest of the support - thermal, mechanical, electrical (power and timing/control/data) all need to be factored in, and controlled to meet the mass specifications. In this task we will undertake the design, and fabrication, of a prototype structure - aimed at a relatively large radius, where the challenge may be greatest, in order to confront all these issues. Such a design must also be appropriate for large scale fabrication, test, and integration.

- **Milestones:**

    – Study group to determine the scope of this activity Year 1

    – Design and build a thermal/mechanical model to demonstrate basic limitations and performance Year 1-3

    – Electrical model version 1 Year 3-5

    – Second version of the electrical prototypes Year 6-8

    – Final production design prototype including assembly and testing process, and methods. Year 8-10

- **Institutes:** SLAC, U Oregon, U Texas Arlington, UC Santa Cruz, BNL, Argonne, Fermilab, LBNL

*2.4.4. Solid State detector task 4: Development of low mass support and cooling structures*

- **Title: Low Mass Support and Cooling for Trackers**

- **Duration:** 10 years

- **Priority:** high

- **Justification:** The previous topic addressed the integration of sensors to local supports and services. Here we focus on the next level, being the global support structure. Depending upon the cooling strategy, the thermal design may emphasize this topic or topic 2 above. The need for cooling has to be understood relative to the duty cycle of the machine. If it cannot be mitigated by the FE power structure, it has to be directly addressed with active cooling. Such cooling, or not, has to be integrated with a low mass global support structure which can also conduct and dissipate heat, as required. The performance of such structures, as in the past, is highly coupled to the available high performance materials, and fabrication method. This is an area of significant experience in the USA but the requirements on mass, scale, and dimensions, go far beyond the current state-of-the-art.

- **Milestones:**

    – Study group to determine scope and specifications Year 1

- Evaluation of possible technologies and development of first prototypes - Years 1-3
- Partial integration test Year 4-5
- Down select among gas, liquid, and passive cooling methods. Year 6
- Large scale design and prototype component fabrication, full scale "sector" demonstration Year 6-10

• **Institutes:** Purdue, Fermilab, LBNL, U Mass, Argonne

*2.4.5. Solid state detector task 5: High efficiency powering and readout schemes*

- **Title: Integration of Novel Electronics Architectures into Detector Modules**
- **Duration:** 10 years
- **Priority:** high
- **Justification:** Trackers for the future electron-positron collider require extremely low mass. The powering and readout infrastructures of the current tracking detectors constitute a large fraction of the material budget, and needs significant reduction to achieve the goals. This task would focus on identifying the appropriate strategies to mitigate power in the front end electronics and the readout. In the case of linear colliders this could also leverage the beam structure when possible. In the case of a circular collider the emphasis would be on high efficiency power conversion methods.
- **Milestones:**
    - Study group to determine scope and specifications Year 1
    - Evaluation of possible technologies and development of first prototypes - Years 2-3
    - Technology down select Year 4
    - System test integration Year 4-5
        - Full scale prototyping in concert with Topic 3 on integration Year 6-8
    - Preparation for production, reliability studies, development of testing infrastructure.
- **Institutes:** SLAC, U Oregon, U Texas Arlington, Argonne, Fermilab

## 3. Muon Detectors & Gaseous Detectors

*3.1. Challenges for Muon Detectors & Gaseous Detectors*

Designs for the next electron-positron collider all propose four different main physics runs at the energy scales for *Z*, *WW*, *ZH*, and *ttH* production, all of which have important final states with high-$p_\mathrm{T}$ muons. To maximize the output of the physics program at the new collider, it is imperative to measure these muons with the highest achievable precision and efficiency in a hermetic subdetector that maximizes muon acceptance. It is also critical to measure the muon tracks with high-precision timing to detect new physics signatures with long-lived particles. To achieve superb precision physics measurement, all proposed accelerators would operate with very high luminosities. For example, the FCC-ee is planned to have instantaneous luminosity of $2 \times 10^{36}$ cm$^{-2}$s$^{-1}$ at the $Z$ peak, where $Z$ bosons will be created at a rate of $\approx$ 100 kHz from the $e^+e^-$ collisions. This will result in an event rate of about 3.4 kHz for high-$p_T$ di-muon events from $Z$ decays. When taking into account the bombardment by muons from decays in hadronic jets, the hit rate in the muon detector will be $\approx$ 10 kHz/cm$^2$ in the hottest forward and backward regions. To reach the high luminosity requirement, the next electron-positron collider will be designed to have a bunch crossing time of 20-25 ns, which necessitates fast muon detectors. In large HEP experiment designs, the muon detectors are the outermost tracking chambers, surrounding the inner tracking and calorimeter systems, and typically covering detection areas of thousands of m$^2$. The challenges of the muon detector design include instrumenting large areas with robust and redundant detectors at low cost, good spatial and temporal resolution with eco-friendly gases, and front-end electronics suitable for streaming output. Consequently, **our highest priority is to develop robust, large-area muon/gaseous detectors with fast timing and high spatial resolution**. In this context, it is also very important to study the operational performance of such detectors with eco-friendly gases.

Based on the past development of gaseous detectors used in HEP experiments, it is well known that a wide range of existing or emerging gaseous detector technologies could be suitable to provide either muon identification and the beam bunch-cross identification (BCID) capabilities or precision muon tracking functionalities or both. Most past experiments used different technologies for precision tracking (in the track bending plane), for the second coordinate measurement (in the non-bending plane), as well as for the BCID. Combining these functions within a single technology has become a hot R&D topic for large gaseous detectors since the detector layout and operations could be greatly simplified and the overall detector construction cost could be significantly lowered.

Another hot R&D topic is the development of micro-pattern gaseous detectors (MPGDs) such as the micro resistive-WELL detector (μRWELL) for the outer layers of the inner tracker and the muon detector. The production of these detectors at very large scale with great reliability in a cost-effective way is yet to be demonstrated. Creating a US-based R&D facility for MPGDs at a National Lab is very important for facilitating this cutting edge MPGD R&D in the US. There is an opportunity to join forces with nuclear physics that is currently pushing for establishing such a facility at Jefferson Lab. For inner tracking systems, the development of low-mass gaseous detectors, such as straw tube chambers, drift chambers, and low-mass MPGDs, is our next highest priority in the R&D program.

Important technical R&D is the development of cost-effective high voltage distribution systems, and a precision alignment system for the muon detector. Detailed design of these systems will highly depend on specific experiment design and detector technology choices. Consequently, the priority for these in the pre-CD0 period are lower compared to the above R&D tasks. Finally, we note that the discussion of calorimeter readouts with gaseous detectors is covered in the calorimeter section of this document.

*3.2. Relevant US expertise in Muon Detectors & Gaseous Detectors*

There is a large community in the US with a long history of muon detector and gaseous detector R&D, construction, and operations in high energy experiments at LEP, the Tevatron, and LHC, as well as in space astrophysics experiments and nuclear experiments. Major successful gaseous detectors developed and built in the US include

- Large multi-wire muon detector for the L3 experiment at LEP;
- Large drift-tube based muon detector for ATLAS experiment at LHC;
- Cathode Strip Chambers (CSCs) as muon detector operating in high-rate region for ATLAS and CMS experiments at the LHC;
- Resistive Micromegas, small Thin-Gap Chambers, and thin-gap RPC for ATLAS;
- Low-mass straw drift tube chambers for ATLAS inner tracker, Phase-1 muon detector upgrade, and for astrophysics balloon experiments (PBAR and SMILE) and space-station AMS experiment;
- GEM detectors and electronics for the CMS muon endcap Phase-2 upgrade;
- Upgrade for the endcap readout plane of the ALICE TPC with quadruple-GEMs;
- Large GEM trackers for the SBS experiment and cylindrical micromegas for the CLAS12 inner tracker in nuclear physics at JLAB;
- Low-mass GEM and μRWELL detectors for central and forward tracking at the electron-ion collider;
- Analog and digital ASIC design and production for the ATLAS muon detector Phase-1 and Phase-2 upgrade;
- Front-end and back-end fast electronics for ATLAS MDT muon detector and inner straw tube tracker readout, and triggering.

*3.3. US Muon Detectors & Gaseous Detectors Institutions*

Institutions responding to the survey and expressing interest in muon detector & gaseous detector development for future $e^+e^-$ colliders are Boston University, Brookhaven National Laboratory, Florida Institute of Technology, Jefferson Lab, Michigan State University, Northeastern University, Tufts University, University of California, Davis, University of California, Irvine, University of Florida, University of Massachusetts, Amherst, University of Michigan, and University of Wisconsin.

*3.4. List of Muon Detectors & Gaseous Detectors R&D tasks*

*3.4.1. MDGD task 1: Robust, large area muon/gaseous detectors with fast timing and high spatial resolution that can be operated with eco-friendly gases*

**Subtask 1:** Large-area precision drift tube based chambers, capable of 3-dimensional tracking and BCID tagging, that can be operated with eco-friendly gases. Considering using aluminum tubes (with 400 μm wall thickness) for large area muon detector.

- **Duration:** 10 years
- **Priority:** high
- **Justification:** This allows exploration of using a single detector technology for 3-dimensional tracking with single tube spatial resolution of 80 μm (1 *cm*) in the bending (non-bending) tracking coordinates and tagging the beam bunch cross ID. Drift tubes are very robust detectors with rate capability up to 12 kHz/$cm^2$ for a one-meter long tube. The US has the infrastructure to build very large muon detectors of this type. But so far it has only been used to measure the precision tracking in the bending plane. With promising time measurement with fast TDCs (with a precision of $O$(100) ps), we could identify long-lived particles (new physics signature), and measure the secondary tracking coordinate along the tube. An algorithm in a high-performance FPGA must be developed so that the detector can have streaming readout and tag the correct BCID without using a trigger system. To achieve these challenging performance goals, significant R&D on eco-gas, tube/chamber configurations, and front-end/back-end electronics must be carried out.
- **Milestones:**
    - Optimization studies of tube parameters such as tube length and wire diameters, wire locator, eco-friendly drift gas, and signal gain. Year 1 - 2.
    - Design and build front-end electronics capable of applying HV and readout at the same end of the tube allowing dual-readout for 3-dimensional tracking. Year 2 - 3.
    - Based on the dual readout results, design a "mean-timer" to measure the non-bending coordinate track position. Year 3 - 4.
    - Read out drift tube signals using a trigger-less streaming mode and use an FPGA to build events for data recording. Year 5 - 6.
    - Build full-size prototype detectors with new electronics and perform cosmic ray and test beam studies to demonstrate the overall performance. Year 7 - 10.
- **Institutes:** UMass Amherst, U. Michigan, UC Irvine, Tufts U.

**Subtask 2:** Thin-gap MPGDs with fast timing for large-area muon detector that can be operated with eco-friendly gases.

- **Duration:** 10 years
- **Priority:** high
- **Justification:** MPGDs with a thin drift gap of 1 mm or less and a single amplification stage promise to achieve nanosecond time resolution, that will allow precise BCID tagging for muons from collisions and rejection of cosmic ray muons. This timing resolution will be an order of magnitude better than the MPGD performance in current experiments. This detector type is particularly suitable for instrumenting the forward and backward directions as the highest rates occur there. The short gap size requires either a higher gas gain than commonly used in state-of-the-art MPGDs or operation with pressurized gas to increase primary ionization. For both approaches, feasibility with eco-friendly gases and robustness, in particular for large detectors operated in this way, must be investigated.

- **Milestones:**
    - Construct medium-size prototypes and characterize performance with existing electronics and eco-friendly gases. Year 1 - 3.
    - Build a full-size MPGD prototype detector with new electronics and perform cosmic ray and test beam studies to demonstrate the performance. Year 4 - 7.
    - Read out signals in streaming mode and use FPGA to build events for data recording. Year 8 - 10.
- **Institutes:** BNL, FIT, JLAB, U. Michigan, U. Wisconsin

**Subtask 3:** Common electronics development for drift tubes and MPGDs

- **Duration:** 10 years
- **Priority:** high
- **Justification:** Testing of new detectors is often hampered by the lack of appropriate frontend and DAQ electronics in sufficient quantities. The MDGD and ASIC groups will collaborate to produce these early on to facilitate the testing of MDGD prototypes.
- **Milestones:**
    - Develop high-resolution TDC/FADC ASICs for timing digitization. Year 1 - 5.
    - Front-end electronics design with time resolution of $\mathcal{O}(100$ ps). Year 5 - 8
    - Develop and implement pattern recognition and segment-finding algorithms inside FPGAs for tagging the BCID and rejecting cosmics. Year 5 - 10.
- **Institutes:** BNL, FIT, JLAB, UMass Amherst, U. Michigan, UC Irvine, U. Wisconsin

*3.4.2. MDGD task 2: US-based R&D facility for MPGDs at a National Lab in collaboration with Nuclear Physics*

- **Duration:** 10 years
- **Priority:** high
- **Justification:** There are over 40 US research institutions involved in MPGD development or activities for experiments in different fields of physics including HEP and NP. They have benefitted from the Micro Pattern Technology (MPT) shop at CERN in the past to produce MPGDs and to perform R&D and optimization on MPGD technologies. However, the global community has been growing swiftly and the MPT currently is the only MPGD source in the world and often struggles to meet demand in a timely fashion. Specifically, there is no such facility in the US to accommodate the future electron-positron collider community's need for MPGDs. The NP community is currently pushing for such a facility and there is an opportunity to join forces with them. Such a facility with similar capabilities as at CERN would be developed at Jefferson Lab to serve the needs of both HEP and NP communities.

- **Milestones:**
  - Development of diamond-like carbon (DLC) foils for large-area resistive MPGDs ($\mu$RWELL and resistive GEMs), which currently can only be produced by a Japanese company. Year 1-3.
  - Fabrication of all elements for complete resistive large-area MPGDs. Year 4-10.
- **Institutes:** FIT, JLAB

*3.4.3. MDGD task 3: Low-mass gaseous detectors for outer region of main tracker*

**Subtask 1:** Develop low-mass straw tubes (with a wall thickness of about 0.04% of a radiation length and made of 75 µm mylar plus 18 µm aluminum) with dE/dx and dN/dx capabilities for inner tracker or for high-eta muon tagger. Since the supporting structure and readout electronics of the straw tubes are only at the ends of the tubes, the radiation length per tube layer will be a factor of 10 smaller compared to a silicon layer. Therefore the straw tube-based tracker can use many layers for track pattern recognition including the identification of long-lived particle decay vertices with high efficiency.

- **Duration:** 10 years
- **Priority:** high
- **Justification:** Allows exploration of 3-dimensional tracking with spatial resolution of 150 µm in the bending plane and 1 cm resolution in the non-bending plane (the 2nd coordinate alone the tube direction). Aiming to tag the beam BCID for high-pT isolated charged tracks within 2-3 bunching cross time (50 - 75 ns). The required R&D for front-end electronics should have large common features. But the front-end board design for the straw inner tracker will need significant optimization highly depending on the straw geometry configurations and the final experiment tracker layout.
- **Milestones:**
  - Develop the straw tube end-plugs, wire locator, and tube grounding method. Study the tube geometry configurations (tube wall thickness, wire diameter, and tube length), and construction method. Year 1 - 3.
  - Design and build two straw tube chambers with 15 and 6 mm tube diameters and with different length up to 1.5 meters long. Year 4 - 5.
  - Design and build the readout electronics, as well as the triggering algorithm implementing in FPGA. Year 6 - 7.

- Perform cosmic ray and test beam tests to demonstrate the 3-dimensional tracking and triggering performance. Year 8 - 9.
- Study the straw tracker particle ID capability through the dE/dx measurement, and combing fast time measurement (with time resolution ≈0.1 ns). Year 10.

- **Institutes:** U. Michigan, Tufts U.

**Subtask 2:** Low-mass MPGDs with 2D readout for tracking

- **Duration:** 10 years
- **Priority:** high
- **Justification:** Planar and cylindrical MPGDs that employ only thin foils in the active area are an option for fast (few ns res.) central and forward trackers with low mass (≈0.4% of a rad. length per layer) and high spatial resolution (≈75µm) in two dimensions. MPGDs can cover large areas in the outer region of the tracker volume in a more cost-effective manner than silicon.
- **Milestones:**
    - Construct medium-size prototypes and characterize with existing electronics. Year 1 - 3.
    - Build full-size prototypes with new electronics and perform cosmic ray and test beam studies to demonstrate the performance. Year 4 - 7.
    - Read out signals in streaming mode and use FPGA to build events for data recording. Year 8 - 10.
- **Institutes:** BNL, FIT, JLAB

**Subtask 3:** Development of a low-mass drift chamber with good tracking and high-precision momentum measurement

- **Duration:** 10 years
- **Priority:** high
- **Justification:** Low mass gaseous detectors provide good tracking and timing. A novel feature of this detector is that it is instrumented with readout electronics implementing the cluster counting/timing techniques, allowing for excellent particle identification over most of the momentum range of interest. The total amount of material in the radial direction is about 1.6% $X_0$, reaching about 5% $X_0$ in most of the forward regions.
- **Milestones:**
    - Construct medium-size prototypes and characterize with existing electronics. Year 1 - 3.
    - Demonstrate the cluster counting method with the prototype detector. Year 3-5.

- Build full-size prototypes with new electronics and perform test beam studies to demonstrate the performance. Year 6 - 10.

- **Institutes:** BNL, FIT

*3.4.4. MDGD task 4: Services and infrastructure for gaseous detectors*

- **Title:** Design and test of HV systems and gas systems
- **Duration:** 8 years
- **Priority:** medium
- **Justification:** Non-commercial services can reduce cost and consequently allow for service designs with high granularity and optimal detector control.
- **Milestones:**
    - Development of high voltage generation and distribution system. MDGD systems typically comprise a large number of modules that need to be powered individually. Years 1-8.
    - Design and build a prototype alignment system for large muon chambers, including optical sensors and a readout system to demonstrate relative alignment of drift tubes with an accuracy of $\approx$ 20 microns. Year 6 - 10.
- **Institutes:** U. Florida, Tufts U., U. Wisconsin

## 4. Calorimeters

The rapid pace of discovery and innovation in collider detectors is largely driven by the quality of the information content in detector data that fuels advanced algorithms and machine learning techniques. Calorimetry for future $e^+e^-$ colliders will have an increasingly central role in the performance of the physics program with major challenges in:

- Suppressing beam-induced backgrounds,
- Maximizing statistical power through Higgs and weak boson decays into jets, and
- High fidelity, high resolution particle-flow reconstruction for low systematic event discrimination and measurement.

Multiple technologies are being pursued for calorimeter options. The gains from co-design of the calorimeter in conjunction with other major collider sub-systems and the foreseen advanced algorithms for event reconstruction are known to be significant. For instance, the precision timing requirements follow from particle-flow algorithm (PFA) and particle identification goals. The roadmap of calorimeter development stresses the need for milestones on front-end performance, verified with test beam, to accurately model and simulate the impact of detector-level choices on the physics program.

The following list is of major calorimeter technologies where significant US contributions are foreseen:

- High granularity silicon sampling calorimeters with embedded CMOS MAPS readout,
- Noble liquid calorimeters, and
- Optical calorimeters: scintillating based sampling and homogeneous calorimeters.

There are also several topical areas co-developed for calorimeter use:

- Calorimeter readout electronics,
- Calorimetry with precision timing, and
- Calorimeter optimization for particle-flow.

A list of R&D topics for calorimetry for Linear Colliders and Future Circular $e^+e^-$ colliders, including scope, schedule and prioritization, has been compiled based on US community-wide feedback and P5 input surveys. Similar compilations are being organized in the context of the ECFA and CERN DRD initiatives, CPAD, and project-specific publications on calorimetry [10].

*4.1. CMOS MAPS Development for Calorimetry*

CMOS Monolithic Active Pixel Sensors (MAPS) provide the potential to develop the next generation of ultralight trackers and highly granular electromagnetic calorimeters for Higgs factory detectors. This technology may achieve the ambitious goals of such detectors, but an R&D effort is needed to reach the required performance. There is much commonality between the requirements for tracking and calorimetry, meaning the effort will be conducted jointly. An example of the potential power of the granularity of this approach is illustrated in Figure 4(a) and Figure 4(b) showing the comparison of EM CAL responses for 13 mm$^2$ and 2.5 x 10$^{-3}$ mm$^2$ granularity.

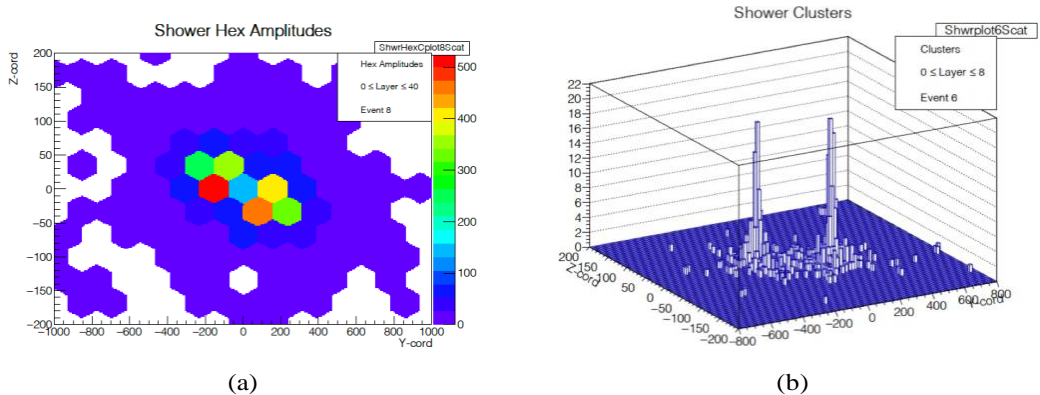

(a)      (b)

Figure 4: Transverse distribution of two 10 GeV showers separated by one cm. LEFT: Pixel amplitudes in the ILC 13 $mm^2$ TDR design. RIGHT: Clusters in the first 5.4 radiation lengths in the new SiD digital MAPS 2.5 x 10$^{-3}$ mm$^2$ design based on a GEANT4 simulation

- **Title:** Development of CMOS MAPS-based Electromagnetic Calorimetry

- **Description:** Development of CMOS MAPS system in common with the tracking development of this application. Current effort will be increased to realize large prototypes, and eventually a multilayer module for beam tests to demonstrate performance.

- **Duration:** 10 years.

- **Milestones:**

  - FY23-24: Develop power and signal distribution schemes compatible for cal and tracking, in addition to evaluating first pixel results.

  - FY25: Design PCBs with variations for the services balcony at the edge of sensors. Submission for sensors for large prototype active layers. Understand options for alternative foundries.

  - FY26: Prototype attachment of sensors to PCB, probably with a conveyor oven so large production is feasible.

  - FY27: Build prototype multilayer section with edge cooling and prepare/begin beam test.

  - FY28: Complete beam tests with technical verification.

  - FY29-32: Design, construct and test MAPS ECal modules based on final design of sensors and sampling layer configuration.

- **Priority:** High

- **Justification:** The design and testing of the SiD ECal based on silicon sensors segmented into 1024 13 mm$^2$ sensors read out by a single chip bump bonded to the sensor (the KPiX ASIC) provides the basis for an excellent linear collider ECal. This concept can be improved in function and reduced cost by replacing the sensors and chips with MAPS. A project has started in this direction [11, 12], but full development and testing remains. A plan for this over the coming years is well coordinated with the timeline of the Higgs factory.

- **Institutes:** SLAC, University of Oregon

*4.2. Noble Liquid Calorimetry*

A highly granular noble-liquid sampling calorimetry was proposed for an electromagnetic calorimeter of an FCC-ee experiment due to its excellent energy resolution, linearity, stability and uniformity. In addition, the noble-liquid calorimetry can be optimized in terms of granularity to allow for 4D imaging, machine learning or - in combination with the tracker measurements - particle-flow reconstruction. This makes it an attractive option for experiment on ILC as well with its longer interbunch time. The radiation hardness of noble-liquid calorimetry makes the R&D investment appealing since it will naturally evolve into a calorimetry solution for the future FCC-hh experiment.

A noble-liquid calorimeter adapted to the central region of an FCC-ee experiment is described in [10], with a cylindrical stack of absorbers, PCB based readout electrodes and active gaps with an inner radius of 2.1 m, to achieve finer longitudinal (12 vs. 4 in AT-LAS) segmentation for PFA. The excellent EM resolution is simulated to be $\sim 8\%/\sqrt{E}$ for LAr calorimeter, while other noble liquid options (LKr, LXe) are being explored as well. Cold electronics is the enabling technology and high priority R&D topic to overcome the cross-talk challenge and achieve superior noise performance ($\sim$5x better SNR than warm electronics).

For this technology to be the basis for a future $e^+e^-$ collider, significant test beam verification has to be achieved in advance of the 2031 milestone for deciding on calorimeter designs as part of a full detector conceptual design report. An international R&D collaboration (ECFA-DRD6) with strong participation of US institutes has been formed to coordinate this effort effectively.

- **Title:** Noble Liquid Calorimetry
- **Description:** Test Beam verification of the noble liquid calorimetry with cold electronics readout. Demonstration of cold electronics performance with PCB based readout electrodes at Phase 1, and construction of full depth calorimeter module for test beam measurements at Phase 2
- **Duration:** Phase 1 (2024-2027), Phase 2 (2028-2033)
- **Priority:** High
- **Justification:** 4D imaging calorimeter with excellent energy resolution, linearity, stability and uniformity for high granularity PFA with reasonable cost and long term impact, significant expertise on cold electronics development and integral system design of noble liquid detectors in US institutes. Close coordination with the task of ASICs for extreme environments in the Readout systems and ASICs group.
- **Milestones:**
    - FY24-27: Cold ASIC development and integration test with PCB based readout electrodes in the cold box
    - FY28-29: Full depth ($\geq 22X_0 : \sim 1.0m \times 0.5m \times 0.5m$) test beam module construction with 64 absorbers and readout electrodes for test beam measurements
    - FY30-33: Full physics performance study of the noble liquid calorimeter module with test beam initial validation for a 2031 CDR and finalized for a 2033 TDR
- **Institutes:** University of Arizona, Brookhaven National Laboratory, Columbia University, Stony Brook University, University of Texas at Austin.

*4.3. Optical Calorimeters: Hybrid Dual-Readout Calorimetry*

Jet energy resolutions of 3-4% for jets with $p_T$ of 50-150 GeV while still maintaining state-of-the-art measurements of electrons/photons has been achieved in full simulation designs of hybrid dual-readout calorimeters [13]. The hybrid method uses segmented homogeneous crystals for the electromagnetic calorimeter and cherenkov/scintillating fibers with time-domain readout for the hadronic compartment. The crystals achieve electromagnetic resolutions of better than $3\%/\sqrt{E}$ with sub-percent constant term, but have low response to hadronic shower energy deposition. The response compensation from reading out the cherenkov light with a separate wavelength-filtered SiPM on the crystals allows the hybrid dual-readout system with the fiber hadron calorimeter to achieve excellent, calorimeter-only jet energy resolution. The particle flow combination of the best measurements from tracking and dual-readout calorimetry show high performance in the accuracy of the jet particle composition and the highest performance PFA jet energy resolution [13].

For this technology to be the basis for a future $e^-e^-$ collider, significant test beam verification has to be achieved in advance of the 2031 milestone for deciding on calorimeter designs as part of a full detector conceptual design report. The validation of segmented crystal with filtered SiPM readout and full-scale combined crystals and fibers are important milestones. The precision timing goals span tens of picoseconds in the front EM section to a hundred picoseconds in the rear fiber readout.

- **Title:** Hybrid Dual-Readout Calorimetry
- **Description:** Test Beam verification of the dual-readout resolution gains and photon/electron resolution. Smaller-scale channel counts at Phase 1, and combined cubic meter scale at Phase2
- **Duration:** Phase 1 (2023-2028), Phase 2 (2029-2033)
- **Priority:** High
- **Justification:** One of the leading, enabling calorimeters for the highest jet and electromagnetic resolutions for high granularity PFA with precision timing. Builds on DOE supported CalVision expert US team with significant international calorimeter collaboration as part of IDEA.
- **Milestones:**
    - FY23-25: Crystal cherenkov signals measured at 50 photoelectrons/GeV effective with separate scintillation readout achieving $3\%/\sqrt{E}$ EM performance
    - FY26-28: Combined crystal and fiber calorimeter readout with hybrid dual-readout multi-signal readout achieving performance goals within limited containment volume
    - FY29-33: Full physics performance cubic meter dual-readout hybrid model with test beam initial validation for a 2031 CDR and finalized for a 2033 TDR
- **Institutes:** Argonne National Laboratory, Fermilab, Oak Ridge National Laboratory, Caltech, University of Maryland, University of Michigan, MIT, Princeton University, Purdue University, Texas Tech University, and University of Virginia.

*4.4. Optical Calorimeters: Scintillator tiles with SiPM Readout*

- **Title:** Hadron Calorimeter Development
- **Description:** The hadron calorimeter is an essential component of the Particle Flow Algorithm approach to achieving the required jet energy resolution for e+e- physics goals. The hadron calorimeter technology must support individual charged particle tracking through the calorimeter, allow detailed imaging of energy depositions for track-shower association and separation of close-by showers, while providing good energy resolution for the direct measurement of the energies of neutral particles. Recently there has also been discussion of the benefits of providing precise timing in calorimeter layers to facilitate the separation of shower components.

- **Duration:**
    - FY24-26, Simulation and optimization of design, including timing
    - FY26-29, Specification of prototype layers, readout, services; beam tests of prototype
    - FY29-31, Mechanical and electrical design of barrel and endcap modules

    R&D Milestones:
    - FY26 - Completion of simulation studies, active layer specification
    - FY28 - Prototype assembled
    - FY29 - Prototype tested
    - FY31 - Barrel and end-cap module designs complete

- **Priority:** High

- **Justification:** The assembly and testing of a large prototype scintillator-based hadron calorimeter module by CALICE has provided many valuable results for hadron calorimetery at a linear collider detector. This technology is also being used for a major upgrade of the CMS end-cap calorimetry - the HGCAL. However, much R&D remains to be carried out in order to be able to specify the technical details for the use of this technology in an e+e- detector.

- **Institutes:** University of Texas at Arlington, Florida State University, Northern Illinois University, University of Maryland, University of Minnesota, SLAC.

*4.5. RPC-based Digital Calorimetry*

- **Title:** Development of RPC-based Digital Calorimetry
- **Duration:** 10 years
- **Priority:** Low
- **Justification:** The concept of the RPC-based digital hadron calorimeter has been validated with various stages of prototyping and testing. Further development of the technology is planned with a low-resistivity glass for increased rate capability of the RPCs, a gas recycling facility and a high voltage distribution system. This is a low-cost technology that can cover large volumes. This technology can be dual-use for extended decay volumes surrounding the calorimeter.
- **Milestones:**

- FY24-26: Development of low resistivity glass. The purpose of this R&D is to develop low resistivity glass with the optimum resistivity to allow larger counting rates but still have the desirable RPC performance.

- FY27-28: Development of high voltage generation and distribution system. A system consisting of a single power supply per module together with a distribution system to the layers needs to be developed.

- FY29-30: Development of a gas recycling facility. For cost reasons and to protect the environment the gas used by larger PC systems must be recycled.

- FY31-33: Prototyping and test beams. Building and commissioning of the final pre-production prototype for finalizing design and performance.

- **Institutes:** University of Iowa, Coe College, Fairfield University, and University of Mississippi

*4.6. US Calorimeter Institutions*

Institutions responding to survey and expressing interest in calorimeter detector development for future $e^+e^-$ colliders: Argonne National Laboratory, Brookhaven National Laboratory, Fermilab, Oak Ridge National Laboratory, SLAC, University of Arizona, Caltech, Coe College, Columbia University, Fairfield University, University of Iowa, University of Maryland, University of Michigan, University of Mississippi, MIT, Northern Illinois University, Northwestern University, University of Notre Dame, University of Oregon, University of Pennsylvania, Princeton University, Purdue University, Stony Brook University, University of Texas at Arlington, University of Texas at Austin, Texas Tech University, and University of Virginia.

## 5. Particle ID

*5.1. Challenges for particle identification at high-energy electron-positron colliders*

An experiment operating at a future electron-positron collider equipped with Particle IDentification (PID) capabilities, in particular with the capabilities of distinguishing between charged hadron species, would enable a compelling physics program. The $e^+e^-$ B-factories and LHCb have demonstrated the importance of hadron identification for precision flavor physics. This capability impacts also $\tau$, top, W, Z, and Higgs physics.

Given the space and material constraints, the implementation of an effective PID detector is challenging, and the physics requirements demand innovative solutions beyond the currently available technologies in order to also maintain state-of-the-art vertexing, momentum resolution, calorimetry, and detector hermeticity. PID is especially important for full exploitation of the physics program of a detector operating at the Z pole.

There are several physics drivers to be considered that are achievable only with an excellent hadron identification:

1. jet flavor tagging (b, c, s, u/d/g)
2. jet charge tagging for asymmetry measurements

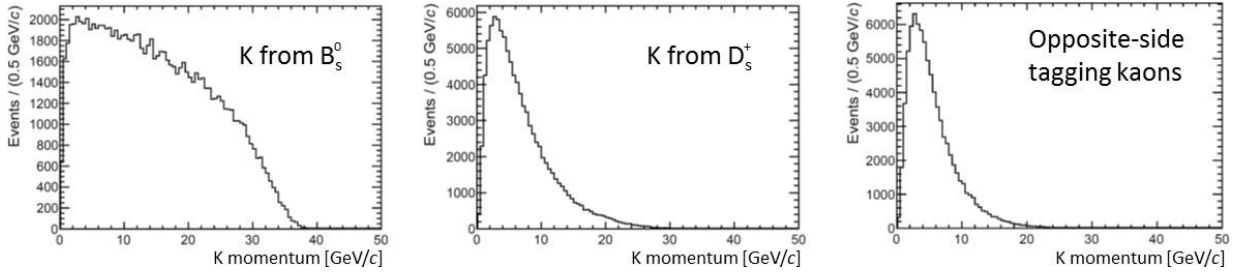

Figure 5: Momentum spectra for kaons occurring in $Z^0$ events containing a $B^0 \to D^{\pm}K^{\mp}$ decay [14].

3. reduction of the combinatorial background with hadron identification and excellent momentum resolution
4. in measurements of CP asymmetries in neutral systems, it is necessary to determine the flavor of the decaying b-hadron, namely whether it was a $B$ or a $\overline{B}$ when it was produced
5. in measurements of rare and forbidden heavy-flavor transitions, the identification of daughter particles in decays with the same topology, for example $B_s \to D_s K$ has the same topology as the prevalent $B \to D_s \pi$

Recent work [15] highlighted the utility of charged kaon particle ID for strange quark tagging and specifically for measuring H $\to \bar{s}s$. The importance of particle ID is also stressed in [14, 16]. Figure 5 illustrates the kaon momentum range for Z pole physics. To satisfy the requirements posed by all these physics goals it is necessary to identify charged hadrons in a momentum range up to approximately 50 GeV.

A single technique is not sufficient to cover the whole momentum range, and the constraints present in different detector concepts lead to different optimizations (see Fig. 6). In the low momentum range the time of flight measurement is a promising technique, which can be pursued with a variety of approaches, for example, silicon-based sensors, with large area micro-channel plate (MCP) PMTs detecting Cherenkov photons from a quartz radiator, or integrated timing measurements in the tracking and especially the calorimetry. These can be complemented at higher momenta by either a compact Ring Imaging Cherenkov (RICH) detector or by gaseous-based tracking using ionization measurements. Existing detector concepts, propose either a drift chamber with cluster counting (IDEA) or a Time Projection Chamber (TPC) with dE/dx measurements in either energy or cluster-counting mode (ILD). Such detectors can also markedly enhance the electron-ID especially inside jets. A possible scenario is the combination of time of flight systems and a Cherenkov detector for detectors that employ a silicon tracker, and a similar time of flight system supplementing the gaseous-based trackers with cluster-counting capabilities of IDEA and ILD.

Beyond the direct application to charged particle hadron identification, timing solutions that are integrated with the calorimetry offer the prospect for the improvement in particle-flow-based jet energy resolution through separation of neutral and charged particles in the calorimeters.

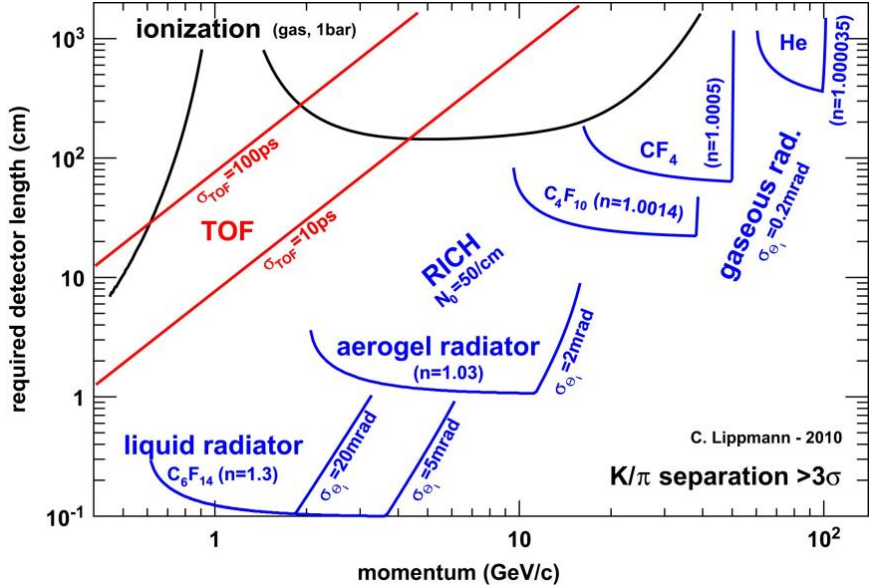

Figure 6: Figure taken from [17]: Approximate minimum detector length required to achieve at least $3\sigma$ K/$\pi$ separation with three different PID techniques. For the energy loss technique we assume a gaseous detector. For the TOF technique, the detector length represents the particle flight path over which the time-of-flight is measured. For the Cherenkov technique only the radiator thickness is given. The thicknesses of an expansion gap and of the readout chambers have to be added.

Specialized time of flight systems, implemented either as wrapper detectors to the tracker or integrated in the electromagnetic calorimeter design, can benefit from a time resolution better than that currently achievable.

### 5.2. Relevant US expertise in particle identification

Large area (O(10) m$^2$) Time-of-Flight (ToF) detectors are being developed for the Electron-Ion Collider (EIC) based on AC-coupled Low Gain Avalanche Diode (LGAD) Silicon sensor technology. They are projected to achieve a timing resolution of around 25–30 ps and a spatial resolution of better than 30 $\mu m$ per single hit, with 1% $X_0$ material budget per detector layer. US institutions involved in this effort include, BNL, FNAL, Los Alamos, Oak Ridge, Ohio State, Purdue, Rice, UC Santa Cruz, and UIC.

The Syracuse University team working at LHCb has significant expertise in Ring Imaging Cherenkov (RICH) Detectors: they were the leading institute in the construction of the CLEOIII/CLEO-c RICH and in the BTeV RICH. Their current interest is in the development of a ps timing/imaging layer to be integrated in the LHCb Upgrade II calorimeter. A promising technology that may break the 10 ps time resolution barrier are the Large Area Picosecond Photon detectors (LAPPD), based on MCP photon detectors. The R&D effort currently ongoing in prototyping timing layers based on LGADs and LAPPDs may evolve in an innovative and cost-effective design for one of the solutions described here.

*5.3. List of PID detectors' R&D tasks*

*5.3.1. LGAD-TOF*

A LGAD TOF detector for a future $e^+e^-$ collider can benefit from the work being done for EIC and LHC. To develop LGAD TOF detectors, the main focuses would be to improve the timing resolution to 10 ps and to make an integrated system design that meets the power consumption, cooling, and material budget requirements. Therefore, the main R&D tasks are to achieve the following goals within 10 years:

*Definition of detector specifications*

- Task duration: 2 years
- Priority: high
- Justification: In order to define the geometrical parameters, sensor and readout technology, mechanical and electrical infrastructure it is necessary to identify some benchmark decays and integrate the proposed detector in one of the proposed detector systems.
    - learn the simulation framework and introduce a simple model of the proposed detector
    - determine the critical specifications to optimize a few benchmark channels (for example, time dependent CP violation golden *B* decay modes).
    - Possible institute: University of Illinois at Chicago, collaborative effort with other institutions.

*Improve timing resolution of the sensor as discussed in Section 2 to 10ps*

- Task duration: 5 years
- Priority: high
- Justification:
    - The best timing resolution of the present LGAD sensor design that has been achieved is around 20 ps. In order to achieve the total 10 ps resolution of the LGAD TOF detector, it is critical to improve the intrinsic timing resolution of the LGAD sensor to below 10 ps.
    - Possible institute: Brookhaven National Laboratory, Fermi National Accelerator Laboratory, University of California, Santa Cruz and University of Illinois at Chicago together with industrial partners.

*Develop low-jitter low power ASIC and front end electronics including power management capability appropriate to different collider bunch structures*

- Task duration: 10 years
- Priority: high

- Justification:
  - A dominant contribution to the timing resolution comes from the jitter of the frontend ASIC and clock distribution system. In order to provide the 10 ps timing resolution of the LGAD TOF detector, it is necessary to keep the jitter contributions below 10 ps.
  - In order to keep the power consumption and material budget under control, it is necessary to develop low power ASIC and electronics with proper power management capability.
  - Possible institute: Brookhaven National Laboratory, Fermi National Accelerator Laboratory, University of California, Santa Cruz and University of Illinois at Chicago together with industrial partners.

*Develop a conceptual, integrated detector system design that meets the power consumption, cooling, and material budget requirements*

- Task duration: 5 years
- Priority: high
- Justification:
  - A conceptual, integrated design of LGAD TOF detector including LGAD sensor, frontend electronics, mechanical support structure and services is needed to demonstrate that such a detector can meet the power consumption, cooling, and material budget requirements at FCC-ee.
  - Such a design will require efforts to develop prototypes of low material mechanical structure with integrated cooling and services.
  - Possible institute: Purdue University together with industrial partners.

*5.3.2. LAPPD-TOF*

Test beam studies on the current generation of LAPPD have demonstrated time resolution of about 20 ps. Ongoing research is focused on improving this performance. Waveform sampling ASICs have demonstrated the capability of achieving 4–6 ps resolution. To achieve the performance needed for this application, the tasks envisaged are:

1. develop the technology to produce devices suitable to cover large detector area in a cost effective manner
2. further adapt the waveform-sampling ASIC to a system involving many channels
3. develop the electronics infrastructure to maintain the performance of a detector type in a large system
4. develop the mechanics solution suitable for a large detector

These goals can be articulated in the following tasks:

*Definition of detector specifications*

- Task duration: 2 years
- Priority: high
- Justification: In order to define the geometrical parameters, sensor and readout technology, mechanical and electrical infrastructure it is necessary to identify some benchmark decays and integrate the proposed detector in one of the proposed detector systems.
    - learn the simulation framework and introduce a simple model of the proposed detector
    - determine the critical specifications to optimize a few benchmark channels (for example, time dependent CP violation golden *B* decay modes).
    - institute: Syracuse, collaborative effort with University of Chicago and industrial partners

*detector element*

- **task duration**: 5 years
- **priority**:high
- **justification**:MCP have a long track record to produce excellent timing resolution (a few ps). They are generally expensive and less suitable to be mass produced. The LAPPD project was a first step towards lowering cost and allowing applications of this technology on large detectors. Vigorous R&D is needed to improve performance to achieve the $\mathcal{O}(1ps)$ goal and prove reproducibility of performance.

*front-end ASIC and front-end PCB*

- **task duration**: 5 years
- **priority:**:high
- **justification:** The ASIC foreseen is based on current technology available to process a small number of channels with few ps time resolution, and the R&D goal is to adapt the concept to a system involving many channels. The ASIC design will be intimately connected with the PCB implementation, where low-mass and tight integration of the components are key design goals because of the limited space available and the need of minimize the overall material budget. An electrical engineer is needed to implement the PCB and work with an industrial partner or collaborators from the ASIC community to implement this design.

*Mechanical integration*

- **task duration**: 4 years
- **priority:**:high

- **justification:** In order to demonstrate the viability of the proposed detector technology, it is important to construct a full size module, including not only sensor and hybrid, but realistic services, additional electronics for on-detector processing and data management, mechanical support and cooling. A mechanical engineer is needed to implement these tasks.

Alternative solutions using gaseous detectors will be studied as well.

*5.4. Ring Imaging Cherenkov Detector*

RICH detectors have been instrumental to the advancement of flavor physics. In order to cover a large momentum range they typically need multiple sections, for example the LHCb hadron identification system relies on two different RICH dectors using gases with two different thresholds[18] and are relatively large detectors. Efforts to develop compact RICH detectors covering a large momentum range are starting[19] and represents an area of R&D that is key to our overall goal and is supported by a strong US expertise.

## 6. Readout systems and ASICs

*6.1. Challenges for readout/ASICs*

Every detector subsystem in the next electron-positron collider will need a dedicated system of electronics to read out detector activity during operation. This requires the design and construction of both on-detector and off-detector electronics. The severe demands for on-detector electronics in collider experiments including spatial constraints, limitations on power dissipation, high data rates, latency requirements, and radiation tolerance typically motivate the use of application specific integrated circuits (ASICs) rather than field programmable gate arrays (FPGAs) or discrete components. As physics performance requirements demand increasingly complex detector systems, R&D in electronics and ASICs is required to accommodate new performance needs.

The main challenges for detector readout for the next electron-positron collider and the corresponding critical areas of research activity fall into the following six categories, which align well with both the Detector R&D Themes (DRDTs) outlined in TF7 (Electronics and On-detector Processing) of the European Committee for Future Accelerators (ECFA) Detector R&D Roadmap [8] and the Priority Research Directions (PRDs) of the 2020 report on the Basic Research Needs (BRN) for High Energy Physics Detector Research & Development [9]. These documents provide detailed justifications and quantitative requirements for each area of research based on the physics goals of the next electron-positron collider. For reference, the qualitative justifications for each area of research, relevant to the next electron-positron collider, are summarized here and task-by-task below:

- **Increased data density:** Physics requirements for high precision spacial, timing and energy measurements at the next electron-positron collider motivate detectors with increased granularity which in turn require electronic readout systems that can manage the related increases in data rate while maintaining low power dissipation and latency.

- **Increased on-detector intelligence including Artificial Intelligence and Machine Learning:** The increased data density requires more intelligent data handling, processing, and selection, as well as on-detector electronics that are closer to the source of data. This need for increased on-detector intelligence (including front-end programmability, advanced data compression, or real-time classification and feature extraction) motivate the incorporation of artificial intelligence (AI) and machine learning (ML) into the electronics.

- **Monolithic sensor ASICs:** Stringent material budgets and granularity for detector sub-systems at the next electron-positron collider lend themselves to a monolithic solution for the sensing element and readout ASIC, which also provides optimal performance and simplified detector design.

- **4D and 5D techniques:** The readout for future 4D/5D tracking detectors and calorimeters will require high performance sampling and excellent precision for measurements of signal amplitude, position, timing, and shape.

- **Emerging technologies:** Detector readout R&D should take full advantage of modern developments in microelectronics including technology with 28 nm feature size and below, 3D/hybrid integration, silicon photonics, open source design and fabrication tools, wireless control and monitoring, and automated design and verification tools.

- **Extreme environments & longevity:** Detectors at the next electron-positron collider will require readout electronics that can accommodate stringent spatial constraints, along with an increased need for fault tolerance and reliability.

*6.2. Relevant US expertise in readout/ASICs*

The US HEP community benefits from an extraordinary amount of experience and institutional knowledge in the areas of electronics design and development. Electrical and ASIC engineers are an essential part of the team and work closely with physicists, creating feasible design specifications that meet the physics goals and then implementing robust systems based on these designs.

Custom ASICs for HEP were first developed in the 1980s at SLAC to read out silicon strip vertex detectors. Since then, the US community has been involved in ASIC design for many other major collider projects, both domestic and abroad. Most recently, US engineering and physicist teams have made significant contributions to the original construction and subsequent upgrades of the ATLAS and CMS experiments. The ongoing HL-LHC upgrade alone involves the design of several custom ASICs in various technologies, including the 65nm complementary metal-oxide-silicon (CMOS) process, that will provide excellent physics performance along with radiation tolerance and longevity required for 10+ years of HL-LHC operation. For the HL-LHC upgrade of the CMS detector, US institutions were responsible for the successful design and delivery of the ECON and ETROC ASICs for the High Granularity Calorimeter and the MIP Timing Detector, respectively. For ATLAS, custom ASICs were produced for several subsystems, namely the Inner Tracker (GBCR, HCCStar, AMACStar), the LAr calorimeter (ALFE, COLUTA), the High Granularity Timing Detector (MuX64), the muon spectrometer (TDC), and for beam monitoring (Calypso).

The DOE national laboratories and US universities each have an essential role in addressing the key R&D challenges. Currently several labs are active in electronics and ASIC design and are able to support tens of scientists, engineers, and technicians in this effort, making them a natural place to generate and coordinate a critical mass of personpower. The labs also benefit from unique infrastructure that can be exploited for broad benefit. US universities also play a critical role, with unique access to on-campus electrical engineering departments and expertise, and have successfully delivered readout electronics projects for collider experiments. Furthermore, universities can naturally provide essential early career personpower to expose and train the next generation of scientists in electronics.

US institutes currently active in electronics/ASICs R&D and anticipated to contribute to future $e^+e^-$ collider efforts include, but are not limited to, Argonne National Laboratory, Brookhaven National Laboratory, Columbia University, Duke University, Fermilab, Lawrence Berkeley National Laboratory, Notre Dame University, SLAC National Accelerator Laboratory, the University of Michigan, and the University of Texas at Austin.

*6.3. List of readout/ASICs R&D tasks*

The following tasks are identified as critical for the effective and timely delivery of suitable readout technology for the next electron-positron collider. These tasks are cross-cutting across the major research challenge areas, and their assigned priorities are compatible with the ECFA Detector R&D Roadmap, the PRDs of the 2020 BRN for HEP Detector R&D, and the priorities of the US groups planning R&D for the next electron-positron collider.

In general, most of the tasks share a common milestone structure related to the iterative design process typical for new electronics and ASICs including (i) conceptual and/or preliminary designs for critical IP blocks, (ii) a prototype chip for evaluating IP block performance, and (iii) specification and design of a prototype ASIC for a specific experiment or task. Considering the timeline for potential projects for the next electron-positron collider, R&D for generic ASIC needs is interleaved with development for experiment-specific designs. Required for success in this iterative process is funding for the full cycle, up to and including fabrication, of approximately three chips (preliminary, pre-prototype, prototype) for each area of development. The associated personpower is anticipated to increase accordingly, such that in the later years of this program, general R&D for new technologies can continue in parallel to designs that account for experiment-specific considerations. As the development of readout naturally aligns with the development of the detectors themselves, and many of the following tasks have synergies among them, funding in this area can facilitate a community enterprise with broad benefits across the R&D effort.

*6.3.1. AI/ML in ASICs*

- **Title:** Build out AI/ML functionality in IP blocks, such as data compression, real-time classification or feature extraction, intelligent power management, or front-end programmability.
- **Duration:** 10 years
- **Priority: High**

- **Justification:** Leverage novel, powerful, and diverse ML-based data handling methods to address the challenges of large complex future collider data-taking, including high input dimensionality (data compression), fast evaluation timescales, and challenging inference tasks (classification, regression, feature extraction).

- **Milestones:**
  - Conceptual design for generic AI/ML in ASICs including simple classification, regression, and compression algorithms in both the digital and/or analog space (FY28)
  - Prototypes for generic AI/ML ASICs respecting expected experimental restrictions on latency, power consumption, granularity, etc. (FY31)
  - Experiment-specific prototypes for AI/ML-based readout ASICs, coherent with trigger and DAQ developments as described in Section 7 (FY33)

*6.3.2. Monolithic sensor ASICs*

- **Title:** Monolithic sensor ASICs includings MAPS, SPADs, and SiPMs
- **Duration:** 10 years
- **Priority: High**
- **Justification:** Ensure the ability of tracking and calorimeter detectors for the next electron-positron collider to read out advanced silicon sensors with the required high granularity and low material budget. Related R&D can address monolithic sensing and readout for several technologies including monolithic active pixel sensors (MAPS), single photon avalanche diodes (SPADs), and silicon photo-multipliers (SiPMs).
- **Milestones:**
  - Conceptual design and development of collaboration and specifications with foundries capable of providing required technology (FY28)
  - Prototypes for evaluating foundry performance (FY31)
  - Experiment-specific prototypes for monolithic sensor ASICs, coherent with sensor development as described in Section 2 (FY33)

*6.3.3. High performance ASICs for 4D/5D detectors*

- **Title:** Electronics for 4D and 5D techniques including multi-function integrated ASICs with high performance analog-digital converter (ADC) or time-to-digital converter (TDC) chips, as well as precision timing.
- **Duration:** 10 years
- **Priority: High**

- **Justification:** Particle ID at the next electron-positron collider will require precise measurements of signal amplitude, shape, and timing across detector subsystems. Calorimeter systems specifically require high dynamic range and minimal jitter for precision timing. This task constitutes the ASIC portion of silicon sensor development articulated in Sections 2 and 5, along with R&D to improve detector-wide clock distribution to maintain timing precision achieved in the readout.

- **Milestones:**
  - Conceptual design for generic 4D/5D IP blocks including phase-locked loops (PLLs), delay-locked loops (DLLs), and ADCs /TDCs (FY28)
  - Prototype with demonstrated $\mathcal{O}$(ps) time resolution (FY31)
  - Experiment-specific prototypes for 4D/5D detector ASICs, in collaboration with detector groups as described in Sections 2, 3, and 4 (FY33)

*6.3.4. IP blocks for 28 nm technology*

- **Title:** Develop general use IP blocks for the 28 nm process with focus on minimal power consumption and high precision.
- **Duration:** 10 years
- **Priority: High**
- **Justification:** Accommodate novel detector challenges such as increased channel density and precision timing, while develop and maintain US experience for this core technology node. Such expertise will allow the HEP community to adapt and migrate to modern foundry methodologies, which is essential as older processes become obsolete and foundries cease production in antiquated technologies.
- **Milestones:**
  - Conceptual design and first prototypes for for general 28 nm IP blocks including PLLs, I/Os, ADC/TDCs, DACs, LDOs, SRAMs, voltage references, etc. (FY28)
  - Second prototype iteration for critical IP blocks demonstrating improved performance relative to first prototypes (FY31)
  - Experiment-specific prototypes in 28 nm (FY33)

*6.3.5. 3D / hybrid integration*

- **Title:** Integrate multiple specialized wafers with various functions into a single monolithic package and incorporate novel wafer stitching strategies to address increased on-sensor demands.
- **Duration:** 10 years
- **Priority: High**

- **Justification:** Fulfill the most stringent performance requirements of future solid state-based detectors as described in Section 2, including resolution, power, and ma- terial budget.

- **Milestones:**
  - Conceptual design and development of collaboration and specifications with foundries capable of providing required technology in multiproject wafers (FY28)
  - First hybrid integrated prototypes to demonstrate performance of technology (FY31)
  - Experiment-specific prototypes making use of 3D / hybrid integration (FY33)

*6.3.6. ASICs for silicon photonics*

- **Title:** Develop ASICs for silicon photonics with high-speed data transmission.
- **Duration:** 10 years
- **Priority: High**
- **Justification:** Keep pace with industry advances in optical transmission standards to achieve very high speed transmission accounting for unique HEP challenges such as distributed data sources.
- **Milestones:**
  - Conceptual design for silicon photonics-based integrated optical modules for > 50 Gbps readout (FY28)
  - Prototypes of critical IP including high-speed serializers, drivers, etc. for evaluating strategies for ASIC and system design (FY31)
  - Experiment-specific prototypes demonstrating silicon photonics ASICs and system integration (FY33)

*6.3.7. Increased data density*

- **Title:** Study electronics/ASIC solutions for challenges associated to high data density, including power & readout efficiency and high date rate systems.
- **Duration:** 10 years
- **Priority: Medium**
- **Justification:** Traditional data storage and processing methods become inadequate to efficiently handle the potentially exabyte-scale datasets anticipated for a trigger-less and/or very high luminosity readout system, requiring innovation to readout systems that can keep pace with advanced accelerators and detectors.
- **Milestones:**

- Conceptual design to reduce expected data rate at ASIC level by order(s) of magnitude (FY28)
- Demonstrate prototype ASIC for high data density on PCB with requisite cooling strategy implemented (FY33)

*6.3.8. Emerging technology*

- **Title:** Incorporate advances in electronics/ASIC technology including open source design/fabrication, wireless control/monitoring, sub-28 nm technology, or automated design/verification.
- **Duration:** 10 years
- **Priority: Medium**
- **Justification:** Keeping pace with emerging technology across various areas, including the industrial sector, can help the HEP community to reduce costs, improve scalability, expedite the design process, minimize errors, and optimize resource allocation.
- **Milestones:**
  - Produce design & prototype for HEP readout ASICs with open source IP blocks and fabrication facilities (FY31)
  - Investigate commercial/industry options for new nanomaterials, Internet of Things (IOT), and/or self-assembly technology in readout concepts (FY33)

*6.3.9. ASICs for extreme environments*

- **Title:** Investigate new approaches to accommodate extreme environments and required longevity, such as reliability & fault tolerance, radiation hardness, or cryogenic temperatures.
- **Duration:** 4 years
- **Priority: Low**
- **Justification:** Given the nature of lepton collider and building on the existing expertise developed in hadron colliders, no electronics R&D is required to deliver readout systems for extreme radiation doses or temperatures. The priority for the next electron-positron collider electronics is in meeting spatial constraints for small or highly granular detectors, as well as fault tolerance and reliability, which must be implemented without violating other power or cooling constraints from the detector, for example via wireless communication. Some other advantages to future $e^+e^-$ detectors could achieved with research in this area, for example cold electronics for noble liquid calorimetry as described in Section 4.

# 7. Triggers and Data Acquisition Systems

## 7.1. Challenges for Trigger and DAQ Systems

The direction for future detectors is to have a higher level of granularity with precise timing information and greater channel capacity, which would result in larger data volumes moved with faster links (The high-granularity electromagnetic calorimeter of ILC/FCC-ee experiments would have 100 times more channels than CMS HGCAL which has 6M) . In addition to challenges in terms of power usage and reliability of the off-detector electronics handling the data operations, intelligent processors closer to the front-end would have to be introduced to handle data selection and reduction to minimize data movement.

Because of the ILC beam structure trigger-less operation is foreseen for the experiments at ILC. However, FCC-ee will operate at the Z pole at higher luminosity, where the event rate is due to the production of Z particles (around 100 kHz), low-angle Bhabha scattering events (around 50 kHz), and the creation of hadrons through photon-photon collisions (around 30 kHz). Creating a conventional hardware trigger system for selecting physics analysis signal events in the FCC-ee's uncontaminated environment should not be overly complicated. However, to achieve the expected accuracy of physics measurements, the trigger system's effectiveness must be known with a precision of $10^{-5}$ at the Z pole (to achieve the physics goals such as the measurements of Z mass and width).

Ultimately, the requirement for extracting physics content at every stage of data acquisition in real-time, at a resolution similar to that of offline processing, will demand the utilization of sophisticated algorithms and hardware. The following R&D areas have been identified to achieve this goal.

## 7.2. Relevant US expertise

As it was outlined in the ASIC/Readout section, the US HEP community possesses extensive experience and institutional knowledge in the field of electronics design and development. Collaborating closely with physicists, electrical and software engineers play a vital role in formulating feasible specifications aligned with physics objectives and implementing robust designs based on those specifications. While the DOE national laboratories naturally serve as hubs for generating and coordinating a critical mass of engineers necessary to address key R&D, the US universities also play a crucial part by leveraging their successful track record in delivering back-end readout and trigger electronics, and DAQ systems with unprecedented network throughput and buffers for previous colliders. Additionally, universities contribute to the exposure and training of early-career scientists in electronics, ensuring the continuity of expertise for future generations.

Notably, recent collaborative efforts between US engineering and physicist teams have yielded significant contributions to the original construction and subsequent upgrades of the LHC Experiment detectors. The ongoing upgrade for the High-Luminosity LHC project involves the design of various customized back-end readout and trigger electronics, employing advanced telecommunications computing architecture (ATCA) standards. These electronics utilize cutting-edge field-programmable gate arrays (FPGAs) and serial optical links capable of operating at speeds up to 25 Gb/s with an overall latency of 12.5 $\mu s$ enabling the inclusion of tracker and high-granularity calorimeter information for the first time. In these systems, improved higher-level object reconstruction and identification, as well as the evaluation of complex global event quantities and correlation variables is planned to be performed. Such evaluations will optimize physics selectivity using sophisticated algorithms that employ particle-flow reconstruction techniques and machine-learning approaches.

Several National Labs and US institutes are presently active in electronics and data acquisition programs within the LHC experiments are: Argonne National Laboratory, Baylor University, Boston University, Brookhaven National Laboratory, Columbia University, Cornell University, Fermilab, Michigan State University, MIT, Northern Illinois University, Princeton, Rice University, Southern Methodist University, Stony Brook, U Penn, UCLA, UM Amherst, University of Arizona, University of California Davis, University of California Irvine, University of Chicago, University of Colorado Boulder, University of Florida, University of Illinois at Chicago, University of Oregon, University of Wisconsin Madison. These Labs and institutions are expected to contribute to future R&D endeavors for $e^+e^-$ colliders. Their involvement ensures the utilization of their expertise and experience in designing advanced electronics systems. The collective knowledge gained from the ATLAS and CMS projects, encompassing the integration of particle-flow reconstruction techniques and machine learning algorithms, will prove invaluable in the development of future $e^+e^-$ collider projects.

## 7.3. Key Areas of R&D

The key areas of R&D addressing the challenged described above and are presented in the following sections. All of the developments will need to keep pace with advancements in FPGA, heterogeneous computing hardware, networking technologies, and online storage systems. Appropriate tools should be developed to leverage these technologies for fast machine learning and DAQ architectures.

### 7.3.1. Application of Machine Learning to TDAQ Systems

Particle physics real-time applications are unique in their requirement for extremely fast inferences, on the scale of sub-microseconds, compared to industrial applications that require longer processing times. The emergence of AI/ML and neuromorphic computing presents a potential opportunity for advancing these applications. Therefore, it is important to invest in R&D to fully understand their potential for future experiments. Such applications can then be used for advance data reduction techniques (based on feature extraction), autonomous operation and calibration.

To integrate a range of real-time ML algorithms based on large datasets into FPGA firmware, tools and expertise are needed. Open-source frameworks like hls4ml have simplified firmware programming and have enabled the integration of advanced AI into high-performance hardware. The ongoing development of ML frameworks can facilitate hardware-software co-design and together with the advances in the processor technologies, can lead to enhancing the sensitivity of future experiments. Such efforts will also enable the training of scientists as experts in data science beyond the field of high-energy physics.

> **Title:** Development of AI, ML, and neuromorphic algorithms and the tools to deploy them for large data volumes and low latency.

**Duration:** 10 years

**Priority:** High

**Justification:** AI/ML/neuromorphic algorithms will be necessary for data reduction, autonomous operation and calibration of the detectors.

**Milestone:**

- Systematically assess and compare the performance, limitations of commercially available high-performance hardware (FPGAs, AI Cores) in relation to AI, ML and neuromorphic algorithms (2024-2033)

- Develop prototype AI, ML, and neuromorphic algorithms that will work with low latency (less than a microsecond) and with the large data volumes expected (2028-2033)

- Develop the tools to deploy them prototype AI, ML, and neuromorphic algorithms on emerging new commodity technology platforms (FPGAs, AI cores, etc.) (2031-2033).

*7.3.2. Achieving High Precision Timing Distribution*

It is crucial to distribute accurate frequency and time references for all readout systems. Traditionally, collider detectors have used machine RF signals as a source of timing references. These clocks are then used to generate timing and synchronization for the detector and are distributed to the back-end electronics through optical fiber links.

The required timing precision will need a precision of 25 ps in $e^+e^-$ colliders and there will be increasing performance demands posed by 4D sensors. In order to properly register events on different detectors, the difference in clock propagation delays must be matched or measured with similar precision. The synchronization requirements may require customized implementation. There are no readily available solutions to this challenge.

**Title: Developing systems with high precision timing synchronization.**

**Duration:** 10 years

**Priority:** Medium

**Justification:** To accurately record events across different detectors, it is crucial to either match or measure the discrepancies in clock propagation delays with a comparable level of precision.

**Milestone:**

- Define system requirements and develop test stands, follow up the developments for the HL-LHC and incorporate the technical achievements and lessons learned (2024-2027)

- Demonstrate 25 ps synchronization across difference distance scales (2028-2033)

*7.3.3. Integration of Modern Computing Hardware*

**Integration of heterogeneous computing hardware to TDAQ architecture**

The TDAQ architecture at the next generation colliders will need to handle enormous amounts of data using software-based triggers that run on various computing resources. To meet the needs, the online processing farm will require next-generation central processing units (CPUs), graphical processing units (GPUs), hybrid CPUs integrated with field-programmable gate arrays (CPU-FPGA), and other commodity processor technolo- gies. Optimal data preparation and distribution using next-generation switching networks, as well as the execution of HEP-specific code or algorithms (including machine learning) on heterogeneous computing platforms, will also be necessary.

**Title:** Integration of heterogeneous computing hardware to TDAQ architecture

**Duration:** 10 years

**Priority:** High

**Justification:** To enable the parallelization of the algorithms acting upon the data at the single event level, heterogeneous computing is necessary.

**Milestone:**

– Continuously monitor and build on upon the the advancements targeting HL-LHC (2024-2029)

– Build demonstrator with commercial hardware and identify the capability limitation (2024-2033)

– Perform system design for reading out at a large scale (2028-2033).

**Streaming design for the trigger DAQ**

An alternative approach to solve the data reduction problem for trigger and data acquisition systems is to operate in a more streaming design, which involves reducing event data at its source and then aggregating and streaming it to downstream computational and storage elements (the data volume is estimated $\sim$ 160 PB/year for an experiment of the $e^+e^-$ colliders.). Further processing and translation of data into higher-level quantities can be performed to achieve the reduction in data throughput and offline computing. Hybrid designs that combine both traditional trigger-based DAQ and streaming-readout are also possible and could simplify DAQ design. This approach should also be investigated in the context of the $e^+e^-$ colliders.

**Title:** Investigations for a streaming design for the Trigger DAQ.

**Duration:** 10 years

**Priority:** High

**Justification:** Streaming design of trigger is an alternative way to address the data reduction problem where the data can be reduced directly at the source (zero suppression) and then aggregated and streamed to storage elements. One notable advantage of this approach is that it eliminates the need for a trigger that relies on custom hardware and firmware. This simplification of the overall system leads to resource savings in both implementation and operation.

**Milestone:**

- Define system requirements and develop prototype a stream-oriented timing system, and data format (2024-2030)
- Define system requirements and develop prototypes for efficient and robust streaming data transfer (2024-2030)
- Define system requirements and develop prototypes for a stream oriented data storage, management and access tools (2024-2030)
- Develop a framework for fully autonomous data acquisition and detector controls systems (2031-2033).

*7.3.4. Improving Data Link Performance and Alternatives*

To accommodate the anticipated increase in data rates from improved granularity and precision timing at future colliders, data links must have improved bandwidth and performance. Improved data links could then read out a larger fraction (or even all of the raw data), which would be beneficial for triggering and event selection. Therefore, different system architectures with massive link capacity must be studied. Besides the targeted developments for future HEP experiments (such as the optical data transmission system development led by CERN), one option is to use COTS optoelectronics with speeds matching 100 Gbps or higher at the back end of the link. However, these must still be compatible with custom front-end designs in areas such as signalling rates, error correction schemes, modulation formats, and protocols.

**Title:** Improving the bandwidth and performance of optical data links

**Duration:** 10 years

**Priority:** Medium

**Justification:** Improvements will enable the continuous streaming of (large fraction or even all of raw) data from the front-end electronics of the detectors for seamless processing.

**Milestones:**

- Participate in the targeted developments in the HEP community, follow up the technical obstacles and solutions (2024-2033)
- Systematically assess the performance, such as power consumption, of commercially available optical links at various speeds, including 40, 50, 100, and 400 Gbps (2024-2033)

- Assess the compatibility and measure the performance of these optical links against the specific requirements the future experiments and custom front-end designs (2028-2033)
- Identify any components that exhibit weaknesses or limitations and work towards improving or replacing them to enhance overall performance (2031-2033).

Another alternative is wireless readout systems, which can have significant advantages over wired ones for some of the high-density detectors. These systems could enable new readout techniques and fast data reduction. However, ensuring RF signal integrity in high-density link environments would require the use of directional antennas, polarization, or attenuating reflections. Detailed design studies are needed to demonstrate the full potential of wireless communication and to create a working wireless readout system on a large scale.

**Title:** Developing high-speed wireless links.

**Duration:** 10 years

**Priority:** Medium

**Justification:** Reducing the number of cables and connectors would have several benefits, including minimizing the presence of dead material within the detectors and simplifying the installation and operation processes.

**Milestone:**

- Investigate commercial chips to construct prototypes (2024-2033)
- Develop custom transceiver chips suitable for high data rate ($>$ 10 Gbps per link) and short distance ($\sim$ 1m) applications (2028-2033)
- Demonstrate a working wireless readout system in large scale (2031-2033).

## 8. Software and Computing

*8.1. Challenges for Software and Computing at the next electron-positron collider*

Software & Computing play a prominent role in the directed R&D, design, prototyping, and building of modern precision collider detectors. They are needed for the broad evaluation and optimization of detector options and their impact on potential physics results. To maximize synergies between individual detector R&D efforts that will meet the physics requirements of the next electron-positron collider, we need to support the development and use of simulation and reconstruction frameworks that accommodate multiple the next electron-positron collider detector concepts. Individual R&D detector studies have to be simulated accurately, but they also need be integrated into and validated in the full experimental context, where the interplay between different detectors plays an important role. Ease-of-use of and engineering support for the software will improve widespread access and enable many contributions to the studies. Well-performing software and infrastructure will allow for efficient and expedient conclusion of studies.

This project should use existing common tools, and develop and contribute to new common solutions in alignment with the international project. International FCC software efforts have been based on the Key4HEP [20] project. Key4HEP is a simulation and reconstruction software framework already used by multiple detector concepts and ILC detector concepts have started to migrate. We believe that consistently using a common software framework is key for the success of the next electron-positron collider detectors and hence it is a cornerstone of this software & computing R&D proposal.

There are many challenges in software & computing and key areas include:

- Underlying community libraries like Geant4 are required for the success of the projects. Their maintenance currently relies primarily on the user base rather than on dedicated developers, and development is needed to simulate novel detectors that push the state-of-the-art. This support needs to be included in the R&D efforts.

- A common basis for developing and executing simulation and reconstruction algorithms enables individual efforts to create a successful detector. Because computing hardware is becoming more heterogeneous, the complexity and intricacy of framework solutions is increasing significantly and can exceed the software capabilities of detector domain experts. Professional support and software engineering contributions to physics and detector simulation and reconstruction development is needed.

- Physics and detector studies require computing resources to produce simulations, to reconstruct simulated detector signals and to analyze the results. R&D efforts need to include computing resources for these tasks. Because these will be distributed and include both high-throughput computing (HTC) and high-performance computing (HPC) facilities, the software will have to support these heterogeneous infrastructures.

- Machine learning (ML) and Artifical Intelligence (AI) will play a prominent role in software and computing and their role is expected to increase in the future. The software and computing infrastructure needs to support AI/ML in a flexible and inclusive way to enable innovation.

Following these challenges, the software and computing R&D part is structured into 4 areas. We focus in our detailed list on tasks where we think the US project could take leadership roles.

1. **Core Software:** includes the community libraries and the core software framework to enable the physics and detector studies. It also includes the support of core software components for the detector R&D efforts.
2. **Infrastructure:** includes the facilities and infrastructure software needed to produce simulations, reconstruct their signals and analyze the output.
3. **Physics Software including AI&ML:** includes engineering contributions to the development of reconstruction algorithms and other physics software.
4. **Coordination:** is tasked to coordinate the different parts of the software and computing R&D activities with national and international projects and activities and to provide oversight.

Here we outline the full set of needs for the next electron-positron collider. In many areas, research topics are also relevant for for the HL-LHC. Common solutions should be pursued.

*8.2. Relevant US expertise in Software and Computing for detectors at future lepton colliders*

As US groups have been leading the design of an ILC detector, SiD, they have also been contributing to the software development. Previously, SLAC led the design and implementation of the simulation software and the reconstruction framework for the SiD detector for the ILC. The LCIO file format and event data model (EDM), which is currently being used by ILC detectors, and which forms the basis for the EDM used by FCC-ee detectors, was developed jointly by researchers at SLAC and DESY.

There is also extensive expertise and experience in Software and Computing for the LHC experiments, much of which is straightforward to translate to the next electron-positron collider. FNAL is the host institute to the US CMS operations program, and ATLAS software and computing has major contributions from ANL, BNL, LBNL, and SLAC. Both experiments are supported through a number of U.S. universities. The sites are generally supported by the Openscience Grid for grid infrastructure software and services, and ESnet for network connectivity. In addition, there have been a number of inter-experiment software institutes performing R&D for HL-LHC and beyond such as HEP-CCE, IRIS-HEP, IAIFI and A3D3.

Historically, the US has made significant contributions to the development of common software packages such as Geant4. Previously, SLAC and FNAL contributed to the development of Geant and more recently Fermilab and ORNL have been developing support for detector simulations for heterogeneous architectures.

*8.3. List of Software and Computing R&D tasks: Core Software*

Core software packages include the simulation and reconstruction framework(s), and the underlying simulation engine that is used by all detector concepts, currently Geant4. These packages incur an ongoing maintenance and education overhead. In addition, we will carry out the following research tasks, which can be grouped under a common heading, but are technically distinct tasks.

*Milestones:*.

- FY24-FY27: Migrate the ILC and FCC detector concepts to Key4HEP and develop features relevant for US detector R&D groups and US HEP priorities, including specific detector descriptions and HPC support. Begin development of GPU-enabled simulation.

- FY28-FY30: Continue community support and establish support for current accelerators (GPUs, FPGAs, etc.) in the framework, evaluate any emerging architectures

- FY31-FY33: Establish support for next-generation accelerators and architectures, evaluate any emerging architectures. Assess and implement any modernization needed for the 2030s.

*8.3.1. Core Software Framework*

- **Title: Support and Evolution of core software framework**

- **Duration:** FY24-FY33
- **Priority:** high
- **Justification:** A common basis for developing and executing simulation and reconstruction algorithms enables individual efforts to create a successful detector. Professional support and software engineering solutions are necessary due to the complexity of current solutions and the heterogeneity of computing hardware.
- **Institutions include:** ANL, FNAL

*8.3.2. Community Simulation Software*

- **Title: Maintenance and Evolution of Community Simulation Software**
- **Duration:** FY24-FY33
- **Priority:** high
- **Justification:** A detailed modeling of detector components and particle interactions in detectors is essential for meeting the physics goals. Underlying the detector simulations are community software packages like Geant4. Successful detector development requires Geant4 and other packages to be maintained and evolved to meet the needs to the detector studies, both in underlying basic software infrastructure and in implementation of physics effects in the simulation.
- **Institutions include:** ANL, FNAL, LBNL, ORNL, SLAC

*8.4. List of Software and Computing R&D tasks: Infrastructure*

Work items to improve the utilization of facilities and not only the compute hardware itself can be grouped under the label "Infrastructure". Detailed milestones in this task depend on the evolution of facilities with respect to provided compute hardware, network connectivity, and available storage solutions.

*Milestones:*.

- FY24-FY27: Establish access to US and worldwide computing and storage resources for the US detector R&D groups. Implement those resources in the detector simulation and physics analysis workflows. Invest in modern analysis approaches like columnar analysis and add analysis using GPUs. Build support infrastructure for large scale AI/ML training workflows.
- FY28-FY30: Maintenance and evolution of workflows to prepare for CD-0. Utilize national efforts for storage like OSG StashCache and the ASCR SuperFacility to improve the efficiency of storage management and access. Enable access to any emerging accelerator concepts.
- FY31-FY33: Maintenance and modernization of workflows to prepare for experiment TDRs

*8.4.1. Resource Provisioning and Workflow Management*

- **Title: Evolution and Operation of Resource Provisioning and Workflow Management**
- **Duration:** FY24-FY33
- **Priority:** medium
- **Justification:** Producing sufficiently large simulation samples is key to design detector components and to optimize whole detector systems. An infrastructure is needed to manage the large amounts of computing resources to produce said simulations. Based on community solutions, this area focuses on providing efficient access to the U.S. computing hardware landscape as well as access to distributed HTC and HPC resources worldwide.
- **Institutions include:** ANL, FNAL

*8.4.2. Storage Management*

- **Title: Evolution and Operation of Storage Management**
- **Duration:** FY24-FY33
- **Priority:** medium
- **Justification:** Sufficiently large simulations for detector and physics studies produce large data volumes that can reach petabytes or beyond. The challenge of organizing the storage of these data samples on a diverse infrastructure of distributed storage facilities falls to storage management. This includes also distributing sub-samples for analysis by the community. Without storage management embedded in community solutions the success of the detector design effort is in jeopardy.
- **Institutions include:** FNAL, SLAC

*8.4.3. Analysis Infrastructure*

- **Title: Evolution and Operation of Analysis Infrastructure**
- **Duration:** FY24-FY33
- **Priority:** high
- **Justification:** The transition from central sample production individual analyses comes with a significant increase in the diversity of software solutions and the number of people needing access to data through storage and computing resources. An efficient and well-supported analysis infrastructure is a key ingredient for timely and detailed detector and physics studies and enables the success of the whole project.
- **Institutions include:** FNAL, MIT

*8.4.4. Large Scale AI/ML Training*

- **Title: Integration, Deployment and Operation of Large Scale AI/ML Training Workflows**
- **Duration:** FY24-FY33
- **Priority:** low
- **Justification:** We expect AI/ML application development and adoption to increase significantly over the next years and play a major role in the final studies for detector designs. Integrating, deploying and operation of large scale training workflows accessing a large volume of data is challenging and cannot be conducted anymore by individual researchers and engineers. This project provides expertise and effort for these large scale AI/ML training workflows.
- **Institutions include:** ORNL

*8.5. List of Software and Computing R&D tasks: Physics Software including AI&ML*

The detector R&D efforts under this umbrella are targeted to meet the physics requirements at the future the next electron-positron collider. Given the ambitious nature of these experiments, detailed understanding of the different R&D efforts with respect to the global physics performance of the overall design is essential for achieving the best possible experimental measurements. This requires developments to the reconstruction algorithms to account for the improving design of the detectors. Additionally, to facilitate the generation of large samples, physics generators can be sped up by improving the use of accelerators and fast simulations can be developed using parametrizations and AI/ML techniques. We can generalize the following milestones.

**Milestones:.**

- FY24-FY27: Engineering support to enable software to run at scale on modern computing infrastructures with GPUs and use AI/ML. Support the consolidation of all reconstruction algorithms into a common core software framework. Support existing solutions and implement new solutions for generative models and fast simulations. Establish a release process for simulation and reconstruction applications and coordinate releases regularly with the domain detector experts and engineers.
- FY28-FY30: Maintenance and development of software for newly emerging accelerators (FPGAs, etc.)
- FY31-FY33: Maintenance and development for the 2040s

*8.5.1. Physics Generators*

- **Title: Optimization and Evolution of Physics Generators**
- **Duration:** FY24-FY33
- **Priority:** medium

- **Justification:** Generator packages are the initial step of the simulation chain and encode the physics underlying the simulation. They are developed by the theory community. Because of the evolution in the hardware landscape towards accelerators (GPUs, FPGAs, etc.) and increased scale and the emergence of AI/ML technique, theorists need engineering support to provide the needed current and future generators for the project.
- **Milestones:**
- **Institutions include:** ANL

*8.5.2. Reconstruction Algorithms*

- **Title: Support Detector R&D domain experts to implement reconstruction algorithms**
- **Duration:** FY24-FY33
- **Priority:** high
- **Justification:** Reconstruction software converts simulated (and later recorded) detector signals back to the particles that produced these signals. Because of the increased complexity of our computing infrastructure, domain detector experts need engineering support to implement reconstruction algorithms efficiently to extract the best possible performance of detectors. This includes both traditional and AI/ML based algorithms.
- **Institutions include:** FNAL

*8.5.3. Simulation*

- **Title: Generative models and fast simulation to accelerate particle and detector level simulations**
- **Duration:** FY24-FY33
- **Priority:** low
- **Justification:** AI/ML techniques to generate particle interactions and fast simulation approaches have the potential to improve the accuracy without sacrificing speed. Their implementation needs engineering support to maintain significant speed-ups compared to their traditional counterparts while keeping their accuracy reasonably high.
- **Institutions include:** FNAL

*8.5.4. Software Release Operations Support*

- **Title: Reconstruction and Monte Carlo Software Release Operations Support**
- **Duration:** FY24-FY33
- **Priority:** medium

- **Justification:** Simulation and reconstruction applications include a variety of algorithms for detector components as well as algorithms that combine information from several/all components to provide a unified reconstructed picture of a particle collision recorded by a detector. To support many different combinations of detector components and many different sub-detector versions including updates to their reconstruction algorithms, infrastructure is needed to collect updates and build consistent releases of simulation and reconstruction applications.
- **Institutions include:** MIT

*8.6. List of Software and Computing R&D tasks: Coordination*

Given the international nature of the software and computing tasks and the diversity of detector R&D concepts that need to be simulated, separate work items for the coordination of US efforts and for the coordination with the international community are warranted. This will establish a leadership role for the US in software and computing for the international experiment, and maintain a coherent effort of the US detector R&D groups.

*8.6.1. International S&C Coordination*

- **Title: International S&C Coordination**
- **Duration:** FY24-FY33
- **Priority:** medium
- **Justification:** Future collider projects and their detectors are international endeavors. They require geographically and monetarily distinct collaborators to work effectively on a common infrastructure and physics product. The US needs to be represented in the efforts to coordinate the detector groups including software and computing to make appropriate contributions and to help shape the direction.
- **Institutions:**

*8.6.2. Coordination of U.S. S&C Activities*

- **Title: Coordination of U.S. S&C Activities**
- **Duration:** FY24-FY33
- **Priority:** medium
- **Justification:** Past experiences with U.S. contributions to major detector projects indicate that the diversity of involved institutes and personnel in the software and computing efforts will be large. This requires coordination and management effort on the U.S. level to match this complexity.
- **Institutions:**

*8.7. List of Software and Computing R&D tasks: User Support*

- **Title: User Support of U.S. R&D Activities**
- **Duration:** FY24-FY33
- **Priority:** low
- **Justification:** To enhance the productivity of US groups in using the software for detector simulation and physics analysis, dedicated person-power would support user requests, and to plan and coordinate tutorials and documentation.
- **Institutions:**

*8.8. Hardware Resources and Collaboration Needs*

The software & computing activities to support detector and physics studies require the availability of sufficient computing hardware. The processing resources are assumed to be acquired opportunistically through the Openscience Grid or HPC allocations at DOE and NSF SuperComputing installations. Access to facilities for analysis is expected to be provided opportunistically initially as well.

The storage space is not provided on an opportunistic basis. We estimate that 3.75 PB of storage will needed for the international project to cover 10 different detector combinations for two accelerator proposals and including a replication factor of 5. An additional 1 PB of storage is required to hold ntuples and other reduced data formats for analysis. For the first years, we will ignore the cost of long-term storage on tape and count on national labs to provide this service.

In summary, we request 5 PB of storage space per year starting in FY24. We anticipate that the yearly storage needs increases will itself increase in FY27 to 10 PB and in FY31 to 20 PB. These estimates all have large error bars of at least a factor three. A more detailed estimate of the storage needs should be made after funding. This should account for the integrated luminosity at the different stages planned for each accelerator concept, including the high statistics needs for TeraZ at the FCC-ee.

In addition, we note that basic collaboration services like a collaboration-wide user authentication and authorization mechanism will be needed. This is usually provided by the host laboratory of an accelerator. A temporary solution may be needed for any collider concept without a host laboratory, for example with a designated host laboratory for the software and computing project.

## 9. Quantum Sensors and Emerging Detector Technologies

*9.1. Challenges for Quantum Sensor and Emerging Detector Technologies*

As an emerging detector technology, there are significant challenges for incorporating quantum sensors into collider experiments. The largest quantum systems to date are those developed for quantum computing and contain on the order of 100 qubits, either superconducting devices or ion traps, contained in one well-contained system. To develop novel systems for collider experiments at-scale will require significant engineering support and device development. This includes, but is not limited to, the development of cryogenic cooling infrastructure compatible with a collider environment, development of sensor fabrication methods and scalable readout systems, studies of the performance of the detectors in high rate, high irradiation, and high magnetic field environments. Given the nascent stage of quantum computing applications the likelihood of quantum systems, as opposed to quantum materials or quantum devices, being of use for Higgs factories is vanishingly small.

Quantum materials and sensors on the other hand are very well-suited for immediate study for implementation into conventional particle physics detectors. For the purpose of this document, quantum materials refers to monolayers, nanowires, quantum dots and other engineered materials. Recent developments in the tuneability and narrow emission bandwidth of quantum dots, for example, open the door to a novel approach to measuring electromagnetic and hadronic energy in scintillator-based calorimeter, with the potential of obtaining a longitudinal tomography of the shower profile within a single monolithic device [21]. Scintillators could be doped with quantum dots to optimize their wavelength emission to match the photodetetors. In general, composite structures combining low-dimensional materials and nanostructures with established detector technologies can offer unprecedented tunability and improvements in detector sensitivity and performance compared to conventional bulk materials. As an example, work function engineering may allow for increased quantum efficiency (QE) with devices being demonstrated by composite photocathodes with coatings of atomically-thin graphene or boron-nitride (BN). Graphene monolayers on photocathodes increase the work function (WF) thus enhancing emissivity, while BN can decrease the WF and increase QE [22, 23]. Different nanowire systems have been proposed as high-efficiency photocathodes owing both to improved geometric emission probability as a result of their large surface to volume ratios as well as their reduced dimensionality. In addition to enhanced sensitivity, low-dimensional materials may also be used to tune the response spectrum by either exploiting resonance effects, e.g. quantum dot size chosen in view of enhanced sensitivtiy to specific wavelength, or using systems that can cover a broad wavelength region such as twisted bi-layer graphene. Although the challenges are large and a significant amount of R&D work is necessary to implement these ideas into working detectors, the potential gains are immense as such emerging quantum sensors have the capability to detect particles with extremely low energy thresholds – far below 1 eV, extremely good position resolution – of the order of tens of nanometers, and excellent time resolution – below 1 ps. Given the long timescale envisioned for this R&D program and the aspirational nature of these emerging technologies, it is important to continue to pursue the specific detector R&D milestones outlined below.

*9.2. Relevant US expertise in Quantum Sensor and Emerging Detector Technology*

Many institutions have a nascent quantum sensing program, building on existing expertise and infrastructure. Most of these efforts are targeting non-collider applications and are most often of limited scale. Existing efforts are also often collaborative multi-disciplinary efforts, bringing together materials scientists, condensed matter physicists and particle physicists. There exists deep expertise in quantum sensing technology in the country. NIST and JPL, for example, are leaders in the development of parametric amplifiers and superconducting nanowire single photon detectors. The HAYSTAC project was the first high energy physics experiment to use a squeezed state receiver [24], a technique which the Dark Matter Radio experiment is taking to a next level [25]. Single crystals such as gallium arsenide are being used to detect optical phonons excited through interactions with dark photons.

A dedicated effort to advance a quantum sensing technology or an emergent quantum material towards a particle physics experiment does not yet exist to the best of our knowledge. Thus, a dedicated initiative to bring quantum sensing and new emerging quantum materials to collider experiments is very timely, given the ongoing efforts in this area and the promise they hold.

*9.3. US Quantum Sensor and Emerging Detector Technology Institutions*

Partial list of institutions with expressed interest in quantum sensor and emerging detector technologies for future colliders: Oak Ridge National Laboratory, Fermi National Accelerator Laboratory, Lawrence Livermore National Laboratory, Lawrence Berkeley National Laboratory, Caltech, NASA Jet Propulsion Laboratory, National Institute of Standards and Technology, Massachusetts Institute of Technology, Argonne National Laboratory, Brown University, University of Maryland, University of Iowa.

*9.4. List of Quantum Sensor and Emerging Detector Technology R&D tasks*

To date two areas have been identified that would benefit from a directed R&D program, Superconducting Nanowire Single Photon detectors and low-dimensional materials, such as monolayers and quantum dots, for photo-detection and calorimetry. The spectrum of quantum sensors is of course much broader, but these studies seem to be better suited for a generic detector R&D program.

*9.4.1. Superconducting Nanowire Detectors*

- **Title:** Superconducting nanowire single photon detector (SNSPD) development and testing for high energy particles
- **Duration:** 5 years
- **Priority:** low
- **Justification:** SNSPDs are a key emerging detector technology with great potential, but requires further understanding of response and implementation in a collider environment.
- **Milestones:** Within the next five years the below list of milestones should be relatively easy to achieve given the ongoing efforts in this area.
    - Characterization of detector response to various high energy particles
    - Demonstration of successful detector operation in high rate and high radiation environment
    - Demonstration of detector performance after high dose irradiation
    - Demonstration of readout capability exceeding 1000 channels on a single sensor
    - Demonstration of single sensors covering area exceeding 5x5 cm$^2$

- **Institutes:** FNAL, Caltech, JPL, MIT, Argonne, NIST.

*9.4.2. Task 2*

- **Title: Low-Dimensional Materials**
- **Duration:** 5 years
- **Priority:** low
- **Justification:** Low-dimensional materials can be embedded in existing materials to tune their response to the specific application and maximize the overall detection efficiency.
- **Milestones:**
  - Demonstration of enhanced performance of photodetectors using two-dimensional materials integrated into the photocathode.
  - Demonstration of scintillators with enhanced performance through the incorporation of quantum dots.
- **Institutes:** Caltech, ORNL, University of Maryland.

## 10. Acknowledgments

This document is made possible by the significant contribution of the U.S. community, their contribution and participation to the discussions following the conclusion of the Snowmass process. We thank them for their invaluable input to the process.

We also thank the readers for taking their time to review this document and offer their constructive feedback. Thanks to Timothy Andeen, Ken Bloom, Joel Butler, Andrei Gristan, Ted Kolberg, Christoph Paus, Paolo Rumerio, Scott Snyder, and Charlie Young.